\documentclass[11pt,a4paper]{article}
\usepackage{amsmath,amssymb,mathrsfs}
\usepackage{subfigure}
\usepackage{indentfirst}
\usepackage{url}
\usepackage{soul}

\usepackage{graphicx}
\usepackage{color}

\usepackage{cite}

\numberwithin{equation}{section}

\allowdisplaybreaks[4]

\begin{document}

\renewcommand{\subfigcapskip}{5pt}

\renewcommand{\theequation}{\thesection.\arabic{equation}}

\renewcommand{\thefootnote}{\fnsymbol{footnote}}
\setcounter{footnote}{0}

\def\thefootnote{\fnsymbol{footnote}}

%%%%%%%%%%%%%%%%%%%%%%%%%%%%%%%%%%%%%%%%%%%%%%%%%%
%%%%%%%%%%%%%%%%%%%%%%%%%%%%%%%%%%%%%%%%%%%%%%%%%%

\hfill{} IPMU18-0110

\hfill{} MPP-2018-140

\hfill{} TU-1065

\hfill{} MIT-CTP/5026

\vspace{1.0 truecm}

\begin{center}

{\textbf{\Large
Enhanced photon coupling of ALP dark matter \\ \vspace{2mm}
adiabatically converted from the QCD axion
}}

\bigskip

\vspace{0.5 truecm}

{
Shu-Yu Ho$^1$\footnote{ho.shu-yu.q5@dc.tohoku.ac.jp}, 
Ken'ichi Saikawa$^2$\footnote{saikawa@mpp.mpg.de}, and 
Fuminobu Takahashi$^{1,3,4}$\footnote{fumi@tohoku.ac.jp}
} \\[5mm]

\begin{tabular}{lc}
&\!\! {$^1$ \em Department of Physics, Tohoku University,}\\
&{\em Sendai, Miyagi 980-8578, Japan}\\[.4em]
&\!\! {$^2$ \em Max-Planck-Institut f\"ur Physik (Werner-Heisenberg-Institut),}\\
&{\em F\"ohringer Ring 6, D-80805 M\"unchen, Germany}\\[.4em]
&\!\! {$^3$ \em Kavli Institute for the Physics and Mathematics of the Universe (Kavli IPMU),}\\
&{\em UTIAS, WPI, The University of Tokyo, Kashiwa, Chiba 277-8568, Japan}\\[.4em]
&\!\! {$^4$ \em Center for Theoretical Physics, Massachusetts Institute of Technology,}\\
&{\em Cambridge, MA 02139, U.S.A.}\\[.4em]
\end{tabular}

\vspace{1.0 truecm}

{\bf Abstract}
\end{center}

\begin{quote}
\hspace{0.5cm}
We revisit the adiabatic conversion between the QCD axion and axion-like particle (ALP) 
at level crossing, which can occur in the early universe as a result of the existence of
a hypothetical mass mixing. This is similar to the Mikheyev-Smirnov-Wolfenstein effect in neutrino oscillations. After refining the 
conditions for the adiabatic conversion to occur, we focus on a scenario where the ALP produced by
the adiabatic conversion of the QCD axion explains the observed dark matter abundance.
Interestingly, we find that the ALP decay constant can be much smaller than the ordinary 
case in which the ALP is produced by the realignment mechanism. As a consequence, 
the ALP-photon coupling is enhanced by a few orders of magnitude, which is advantageous 
for the future ALP and axion-search experiments using the ALP-photon coupling. 
\end{quote}

\thispagestyle{empty}

\newpage

\tableofcontents

\renewcommand{\thepage}{\arabic{page}}
\renewcommand{\thefootnote}{\arabic{footnote}}
\setcounter{footnote}{0}

%=============================================
\newcommand{\Eq}{&=&}
\newcommand*{\aboxed}[2]{%
  \rlap{\boxed{#1#2}}%
  \phantom{\hskip\fboxrule\hskip\fboxsep #1}&\phantom{#2}%
}
%=============================================

%%%%%%%%%%%%%%%%%%%%%%%%%%%%%%%%%%%%%%%%%%%%%%%%%%
\section{Introduction}
\label{sec:Introduction}
\setcounter{equation}{0}
%%%%%%%%%%%%%%%%%%%%%%%%%%%%%%%%%%%%%%%%%%%%%%%%%%

Dark matter (DM) is one of the outstanding mysteries in astronomy, cosmology, and particle physics.
Since its nature is hardly explained within the framework of the standard model of particle physics,
it is highly expected that there exists some new physics beyond the standard model that provides a solution 
to the DM puzzle. Among various possibilities proposed in the literature, the axion and more generally 
axion-like particles (ALPs) remain the best-motivated candidates for DM.

The axion is a hypothetical particle, whose existence was originally postulated in the context of the strong CP problem of quantum chromodynamics (QCD). 
The Lagrangian of QCD includes a term that violates CP symmetry, whose effect is characterized by a dimensionless parameter $\theta_{\rm QCD}{}^{}$.
Since CP is not a symmetry of nature, the magnitude of this parameter is expected to be an order of unity.
However, measurements of the neutron electric dipole moment limit $|\theta_{\rm QCD}| \lesssim 10^{-10}$~\cite{Graner:2016ses}.
Such a small ${}^{}{}^{}\theta_{\rm QCD}$ is unnatural and calls for an explanation of why QCD preserves CP symmetry to very high accuracy.

One of the  attractive solutions to the strong CP problem is the Peccei-Quinn (PQ) mechanism~\cite{Peccei:1977hh,Peccei:1977ur}.
In this mechanism, one introduces a spontaneously broken global axial U(1) symmetry (called the PQ symmetry) such that
the $\theta_{\rm QCD}$ parameter is replaced with a dynamical variable.
This variable is identified as the axion (or hereafter, the QCD axion), which is a pseudo-Nambu-Goldstone boson associated with a spontaneous breakdown of the PQ symmetry~\cite{Weinberg:1977ma,Wilczek:1977pj}. 
It acquires a potential due to non-perturbative effects of QCD \cite{tHooft:1976ripm,Witten:1979vv,Veneziano:1979ec} and settles down at a CP-conserving point, solving the strong CP problem.
Furthermore, a coherent oscillation of the axion field settling down to the minimum of the potential can account for DM~\cite{Preskill:1982cy,Abbott:1982af,Dine:1982ah}.

The notion of the QCD axion can be straightforwardly generalized to the case of many ALPs, which may appear as low energy consequences of
some fundamental theory such as string theory~\cite{Arvanitaki:2009fg,Acharya:2010zx,Cicoli:2012sz}.
The ALPs have similar properties to the QCD axion but do not necessarily interact with QCD gluons.
Furthermore, they do not  have any particular relation between their mass and decay constant, 
which opens up a wide range of possibilities for searching them in laboratories and astrophysical phenomena.
It should also be noted that the dynamics of the ALP field can account for DM in a similar way to the QCD axion~\cite{Cadamuro:2011fd,Arias:2012az}.
See e.g. \cite{Kim:2008hd,Wantz:2009it,Ringwald:2012hr,Kawasaki:2013ae,Marsh:2015xka,Irastorza:2018dyq} for recent reviews
on the theory, phenomenology, and experimental searches for the QCD axion and ALPs.

In most of the previous works on the QCD axion DM or ALP DM, 
the cosmological dynamics of the QCD axion and ALPs was considered separately as if there were no connection between them.
However, there is no particular reason to exclude the possibility that both the QCD axion and ALPs co-exist 
and that there is some non-trivial interplay between them. Indeed, in the context of the string
axiverse~\cite{Arvanitaki:2009fg,Acharya:2010zx,Cicoli:2012sz} or axion landscape~\cite{Higaki:2014pja,Higaki:2014mwa}, there appear many
axions which generally have non-zero mixings among them.

In this paper, we revisit the cosmological evolution of the QCD axion and an ALP 
when they have a nonzero mass mixing. Since the QCD axion mass vanishes at high temperatures and becomes nonzero around the QCD phase transition, the mass eigenvalues could exhibit a peculiar
behavior due to the mass mixing. 
One of the striking phenomena is level crossing between the axion mass eigenvalues, which happens
if the ALP mass is lighter than the QCD axion mass at the zero temperature, and if the ALP decay constant is
smaller than the QCD axion decay constant.
In this case, the adiabatic conversion of the QCD axion into the ALP (and vice versa) could take 
place~\cite{Hill:1988bu,Kitajima:2014xla,Daido:2015bva,Daido:2015cba},
which is analogous to the  Mikheyev-Smirnov-Wolfenstein effect in neutrino 
physics~\cite{Wolfenstein:1977ue,Mikheev:1986gs,Mikheev:1986wj}. Interestingly, the QCD axion abundance
can be reduced by the mass ratio when the adiabatic conversion takes place.
The adiabatic conversion between the QCD axion and ALP and the related topics such as 
the suppression of isocurvature perturbations and the formation of topological defects
were studied in Refs.\cite{Kitajima:2014xla,Daido:2015bva,Daido:2015cba}, 
but its experimental implications, as well as the precise condition for the adiabatic conversion to take place, 
were not fully explored. We will study these aspects of the adiabatic conversion in this paper.
On the other hand, even if the level crossing does not take place, the mass eigenvalues still evolve in a non-trivial way in a certain case, 
and the final axion abundances can be similarly suppressed compared to the case without mass mixing. 
We will study this case and clarify the conditions for such a non-trivial time evolution of the mass eigenstates to take place. To our knowledge, the latter case was not studied in the literature.
 
The purpose of this paper is threefold. First, we refine the condition of the adiabatic conversion, by taking
account of another relevant timescale which was missed 
in Refs.~\cite{Hill:1988bu,Kitajima:2014xla,Daido:2015bva,Daido:2015cba}. 
We numerically check the validity of the refined adiabatic condition.
Secondly, we study the case in which the mass eigenvalues evolve in a non-trivial way even though
the level crossing does not take place. As we shall see later, the axion abundances are significantly modified in this case.
Thirdly, we explore the experimental implications of the axions with mass mixing.
Specifically, we focus on a scenario where the observed DM abundance is explained due to the adiabatic conversion
(or non-trivial time evolution) of the QCD axion and ALP. In the parameter space of our interest, the ALP (or more precisely, the lighter mass eigenstate to be defined in the later section) gives the main contribution to the observed DM.
In this case, the ALP abundance can be enhanced compared to the case without the mass mixing, which enables the ALP with a relatively small decay constant to be the main component of DM. This is advantageous for future axion search experiments. 

The rest of this paper is organized as follows. In Sec.~\ref{sec:level_crossing}, we briefly review the QCD axion and ALP, and classify their evolution according to the mass eigenvalues and mass eigenstates. In particular, we clarify when the level crossing takes place. 
In Sec.~\ref{sec:cosmology}, we study cosmological evolutions of the QCD axion and ALP to
estimate their relic abundance.
In Sec.~\ref{sec:axion_photons}, we show the viable parameter space in the plane of the axion mass and coupling
to photons and discuss its implications for the future axion search experiments. 
The last section is devoted to discussion and conclusions. 

%%%%%%%%%%%%%%%%%%%%%%%%%%%%%%%%%%%%%%%%%%%%%%%%%%
\section{Level crossing between the QCD axion and ALP}
\label{sec:level_crossing}
\setcounter{equation}{0}
%%%%%%%%%%%%%%%%%%%%%%%%%%%%%%%%%%%%%%%%%%%%%%%%%%

In this section, we describe some basic properties of the QCD axion and ALP(s). First, we briefly summarize the known properties of the QCD axion DM in Sec.~\ref{sec:QCD_axion_DM}. We then introduce ALPs and discuss their similarities and differences to the QCD axion in Sec.~\ref{sec:ALP_DM}. The effect of the possible mass mixing between the QCD axion and ALP is discussed in Sec.~\ref{sec:mass_mixing}. In particular, we define the level crossing between the QCD axion and ALP and clarify its peculiarity.

%%%%%%%%%%%%%%%%%%%%%%%%%%%%%%%%%%%%%%%%%%%%%%%%%%
\subsection{QCD axion DM}
\label{sec:QCD_axion_DM}
%%%%%%%%%%%%%%%%%%%%%%%%%%%%%%%%%%%%%%%%%%%%%%%%%%

The QCD axion, $a^{}$,  is a pseudo-Nambu-Goldstone boson associated with the spontaneous breakdown of global axial U(1) PQ symmetry. At energies below the scale of the PQ symmetry breaking and above that of the QCD phase transition,
it couples to gluons through the following effective interaction,
\begin{eqnarray}
\mathcal{L}_{agg} 
\,=\,
-\frac{\alpha_s}{8\pi} 
\frac{a}{f_a}^{}
G^{b}_{\mu\nu}\widetilde{G}^{b\mu\nu}
~, \label{axion_gluon_coupling}
\end{eqnarray}
where ${}^{}\alpha_s$ denotes the strong fine structure constant, $f_a$ the axion decay constant,
and $G^b_{\mu\nu}$ and $\widetilde{G}^b_{\mu\nu}$ a field strength of the gluon field and its dual, respectively.
Due to the existence of the interaction~\eqref{axion_gluon_coupling}, topological fluctuations of the gluon fields in QCD induce the following effective potential for the axion field,
\begin{eqnarray}
V_{\rm QCD}(a) 
\,=\, 
\chi(T)
\scalebox{1.1}{\bigg[}\,
1-\cos\bigg(\frac{a}{f_a}\bigg)
\scalebox{1.1}{\bigg]}
~, \label{V_QCD}
\end{eqnarray}
where $\chi(T)$ is the topological susceptibility, which depends on the temperature ${}^{}T$ of background radiations. In particular, $\chi(T)$ takes some finite value at low temperatures, while it goes to zero at temperatures much higher than the QCD confinement scale. At low temperatures, the potential has a minimum at $\theta_{\rm QCD} = \langle a\rangle/f_a = 0{}^{}$, which provides a dynamical solution to the strong CP problem, where $\langle a\rangle$ is the vacuum expectation value of the axion field.

The effective potential~\eqref{V_QCD} induces the mass of the QCD axion,
\begin{eqnarray}
m^2_a(T) \,=\, 
\frac{\chi(T)}{f_a^2}
~,
\end{eqnarray}
which depends on the temperature. Here we model the temperature dependence of the QCD axion mass as a power-law function\,:
\begin{eqnarray}\label{axionmass}
m_a(T) 
\,\simeq\, 
\begin{cases}
\displaystyle
\,\frac{\sqrt{\chi_0}}{f_a}
\scalebox{1.1}{\bigg(}
\frac{T_{\rm QCD}}{T}
\scalebox{1.1}{\bigg)}^{\hspace{-0.12cm}n}
& T \,>\,  T_{\rm QCD} 
\\[0.3cm]
\displaystyle
\,\,m_a \equiv m_a(T\to 0)
& T \,<\, T_{\rm QCD}
\end{cases} 
~,
\end{eqnarray}
where $\chi_0$ is computed by lattice QCD and the parameters $T_{\rm QCD}$ and $n$ are chosen such that 
they reproduce the correct magnitude of $m_a(T)$ and that
the value of $m_a(T)$ is matched with the zero temperature value $m_a$ at $T= T_{\rm QCD}{}^{}$. A recent detailed analysis gives the following numerical result on the zero temperature mass~\cite{diCortona:2015ldu}\,:
\begin{eqnarray}
m_a 
=\, 
5.70(7)\,\mu\mathrm{eV}
\scalebox{1.1}{\bigg(}
\frac{10^{12}\,\mathrm{GeV}}{f_a}
\scalebox{1.1}{\bigg)}
~.
\end{eqnarray}
On the other hand, in order to know the temperature dependence of ${}^{}m_a(T)$ at high temperatures, 
it is necessary to analyze non-perturbative effects in QCD. In this paper, we adopt a value of ${}^{}n = 4.08$
based on one of the latest lattice QCD result~\cite{Borsanyi:2016ksw}.\footnote{Recently, the temperature dependence of the axion mass has been directly investigated in lattice QCD by several groups. 
Recent calculations by other groups~\cite{Berkowitz:2015aua,Petreczky:2016vrs,Frison:2016vuc,Taniguchi:2016tjc} 
found similar behavior. (See however Ref.~\cite{Bonati:2015vqz} which obtained a different result.)} 

The QCD axion is produced in the early universe via the realignment mechanism~\cite{Preskill:1982cy,Abbott:1982af,Dine:1982ah}. If the axion field has an initial value, $a_0 =  f_a\theta_0{}^{}$, at the very early stage of the universe,\footnote{Here we assume that the axion field has  a spatially uniform value $f_a \theta_0$ throughout the observable universe. This is guaranteed if the PQ symmetry was broken before inflation and never restored afterward, but it does not hold if the PQ symmetry was restored and got spontaneously broken after inflation. In the latter case, we must take account of the effect of the collapse of strings and domain walls rather than the realignment mechanism~\cite{Hiramatsu:2012gg,Kawasaki:2014sqa,Fleury:2015aca,Klaer:2017ond,Gorghetto:2018myk,Kawasaki:2018bzv}.}
it starts to oscillate around the minimum of the potential~\eqref{V_QCD} when its mass $m_a(T)$ becomes comparable to the cosmic expansion rate $H(T){}^{}$. Such a coherently oscillating axion field can serve as cold DM in the universe. The relic abundance of the QCD axion 
DM,  $\Omega_a{}^{}$,  from the realignment mechanism in the regime $|\theta_0|\ll \pi$ reads~\cite{Ballesteros:2016xej}\footnote{
When the initial angle $\theta_0$ becomes sufficiently large, the anharmonic effect which leads to the enhancement of the axion abundance must be taken into account~\cite{Kobayashi:2013nva}. On the other hand, the value of $\theta_0$ has to be sufficiently small if $f_a$ takes larger values in order to avoid the overproduction of the axion DM.
We note that this fine-tuning of ${}^{}\theta_0$ can be greatly relaxed for low-scale inflation with the Hubble parameter comparable to or less than the QCD scale~\cite{Graham:2018jyp,Guth:2018hsa}.}
\begin{eqnarray}
\Omega_a h^2 
\,\simeq\, 
0.14 \,\theta_0^2 
\displaystyle 
\scalebox{1.1}{\bigg(}
\frac{f_a}{10^{12}\,\mathrm{GeV}}
\scalebox{1.1}{\bigg)}^{\hspace{-0.15cm}1.17} 
~,
\label{QCD_axion_realignment_abundance}
\end{eqnarray}
where $h$ is the normalized Hubble constant, and this formula corresponds to the case in which the QCD axion begins to oscillate before $m_a(T)$ reaches the zero temperature value.
This assumption holds for $f_a \lesssim 3 \times 10^{17}\,\mathrm{GeV}$.

The interactions of the QCD axion with the standard model particles are suppressed by the decay constant $f_a^{}$.
The lower bound on $f_a \gtrsim 4 \times 10^8$\,GeV guarantees the stability of the QCD axion DM on a cosmological time scale~\cite{Mayle:1987as,Raffelt:1987yt,Turner:1987by}. The coupling of the QCD axion to photons is given by
\begin{eqnarray}
\mathcal{L}_{a\gamma\gamma}
\,=\, 
-
\frac{g_{a\gamma\gamma}}{4}  {}^{}
a  {}^{} F_{\mu\nu}\widetilde{F}^{\mu\nu}
~,  \label{axion_photon_coupling}
\end{eqnarray}
where ${}^{}F_{\mu\nu}$ is the electromagnetic field strength tensor and ${}^{}\widetilde{F}_{\mu\nu}$ its dual.
The coupling coefficient ${}^{}g_{a\gamma\gamma}$ reads~\cite{diCortona:2015ldu}
\begin{eqnarray}
g_{a\gamma\gamma} 
\,=\, 
\frac{\alpha}{2\pi f_a}  {}^{} C_{a\gamma}
~, \quad 
C_{a\gamma} 
\,=\,
-1.92(4) + \frac{\cal E}{\cal N}
~, \label{g_a_gamma}
\end{eqnarray}
where ${}^{}\alpha{}^{}$ is the electromagnetic fine-structure constant. The coefficient in Eq.~\eqref{g_a_gamma} contains a model-independent contribution from the mixing with mesons (the first term in $C_{a\gamma}$) as well as a model-dependent contribution (the second term in $C_{a\gamma}$) given by the electromagnetic anomaly ${}^{}\mathcal{E}{}^{}$ and the color anomaly ${}^{}\mathcal{N}$ of the PQ symmetry. The existence of the photon coupling provides a promising way for various direct detection experiments of the QCD axion DM~\cite{Irastorza:2018dyq}.

%%%%%%%%%%%%%%%%%%%%%%%%%%%%%%%%%%%%%%%%%%%%%%%%%%
\subsection{ALP DM}
\label{sec:ALP_DM}
%%%%%%%%%%%%%%%%%%%%%%%%%%%%%%%%%%%%%%%%%%%%%%%%%%

The considerations on the QCD axion described in the previous subsection can be generalized to the case 
where there exist multiple axion-like fields, $\varphi_j{}^{}$, in the low energy effective theory.
The existence of such multiple axion-like fields is considered to be a generic consequence of scenarios 
proposed in the context of the string axiverse~\cite{Arvanitaki:2009fg,Acharya:2010zx,Cicoli:2012sz} or axion landscape~\cite{Higaki:2014pja,Higaki:2014mwa}.
In particular, we can allow for their couplings to gluons and photons,
\begin{eqnarray}
\mathcal{L}_{\rm gluons} 
\,=\,- \left(\,\,\sum_{j=1}^{n_{\rm ax}} C_{jg} \frac{\alpha_s}{8\pi} 
 \frac{\varphi_j}{f_{\varphi_j}} \right) G^b_{\mu\nu} \widetilde{G}^{b\mu\nu} 
~,\quad
\mathcal{L}_{\rm photons} 
\,=\,
- \left(\,\,\sum_{j=1}^{n_{\rm ax}} C_{j\gamma}  \frac{\alpha}{8\pi} 
\frac{\varphi_j}{f_{\varphi_j}}\right) F_{\mu\nu} \widetilde{F}^{\mu\nu} 
~,
\label{general_couplings}
\\[0.01cm]\nonumber
\end{eqnarray}
where ${}^{}C_{jg}$ and ${}^{}C_{j\gamma}$ are model-dependent constants of $\varphi_j{}^{}$, and  ${}^{}n_\text{ax}$ is the number of axion-like fields. Then, it is possible to define a field, $a/f_a \equiv \sum_j C_{jg} {}^{} \varphi_j/f_{\varphi_j}^{}$,
which directly couples to gluons, while the rest of the fields orthogonal to ${}^{}a{}^{}$ do not couple to gluons but  may still have interactions with photons.
In this case, we loosely refer to the former as the QCD axion and the latter as ALPs. Note however that, as we shall consider shortly, 
they are not necessarily the mass eigenstates, in general. 
We hereafter consider the case with $^{}n_\text{ax} = 2^{}$ where we have the QCD axion $^{}a^{}$ and
the ALP $\varphi^{}$, and $\varphi$ has no coupling to gluons. 

In contrast to the QCD axion, ALPs do not acquire a mass from the QCD effects. 
Instead, they may acquire a mass from high energy physics, such as effects of some hidden confining gauge interactions \cite{Kitajima:2014xla}.
Here we focus on their low energy phenomenology and simply treat their mass $^{}m_{\varphi}$ as an arbitrary constant.

The ALP-photon coupling can be written as
\begin{eqnarray}
\mathcal{L}_{\varphi\gamma\gamma} 
\,=\, 
-
\frac{g_{\varphi\gamma\gamma}}{4}^{}
\varphi^{} F_{\mu\nu} \widetilde{F}^{\mu\nu} 
~, \quad 
g_{\varphi\gamma\gamma} 
\,=\,
 \frac{\alpha}{2\pi f_{\varphi}}^{}  C_{\varphi\gamma} ~,
\label{ALP_photon_coupling}
\end{eqnarray}
where ${}^{}f_{\varphi}$ is the ALP decay constant, and ${}^{}C_{\varphi\gamma}$ is a model-dependent coefficient.
Note that the ALP-photon coupling contains the model-dependent contribution only, 
which should be contrasted to the QCD axion-photon coupling~\eqref{g_a_gamma} which includes the model-independent contribution.

The ALP is also produced in the early universe via the realignment mechanism and 
it can account for cold DM for certain values of the parameters~\cite{Cadamuro:2011fd,Arias:2012az}.
Let us assume that the ALP field $\varphi$ is a mass eigenstate with an eigenvalue $m_{\varphi}{}^{}$, which is orthogonal to 
the state corresponding to the QCD axion.
Then, the ALP abundance is estimated as~\cite{Arias:2012az}
\begin{eqnarray}\label{ALPabundance}
\Omega_{\varphi}h^2 
\,\simeq\,
0.3\, \theta_{\varphi,0}^2 
\bigg(\frac{m_{\varphi}}{1\,\mathrm{eV}}\bigg)^{\hspace{-0.12 cm}1/2}
\scalebox{1.1}{\bigg(}
\frac{f_{\varphi}}{10^{12}\,\mathrm{GeV}}\scalebox{1.1}{\bigg)}^{\hspace{-0.14cm}2} ~,
\end{eqnarray}
where $\,\theta_{\varphi,0} \equiv \varphi_0/f_{\varphi}$ is an initial misalignment angle of the ALP field.
Note that, in contrast to the QCD axion, the ALP mass is assumed to be constant in temperature and independent of the decay constant $f_{\varphi}{}^{}$,
which leads to the different powers of the decay constants in Eqs.~\eqref{QCD_axion_realignment_abundance} and \eqref{ALPabundance}.

%%%%%%%%%%%%%%%%%%%%%%%%%%%%%%%%%%%%%%%%%%%%%%%%%%
\subsection{Mass mixing between the QCD axion and ALP}
\label{sec:mass_mixing}
%%%%%%%%%%%%%%%%%%%%%%%%%%%%%%%%%%%%%%%%%%%%%%%%%%

In the previous subsection, we have assumed that
the ALP field ${}^{}{}^{}\varphi{}^{}$ is a mass eigenstate  orthogonal to the QCD axion $a{}^{}$.  However, in light of the general consideration mentioned below Eq.~\eqref{general_couplings},
it is not unreasonable to expect that the two fields ${}^{}a$ and $\varphi$ are not orthogonal to each other,
and that effects of new physics might induce a potential for a linear combination of ${}^{}a$ and $\varphi{}^{}$.
As a concrete example, let us consider the following potential \cite{Kitajima:2014xla,Daido:2015bva,Daido:2015cba}, 
\begin{eqnarray}
V_{\rm mix}(a,\varphi) 
\,=\, 
m_{\varphi}^2 {}^{}  f_{\varphi}^2
\scalebox{1.1}{\bigg[}\,
1-\cos\bigg(\frac{a}{f_a} + \frac{\varphi}{f_{\varphi}}\bigg)
\scalebox{1.1}{\bigg]} \label{V_mix}
\label{eq:mixV}
\end{eqnarray}
in addition to the QCD potential~\eqref{V_QCD}.
The above potential induces a mass mixing between the QCD axion and ALP, which causes several non-trivial effects as discussed below.
Note that $m_\varphi$ is just a parameter and does not necessarily coincide with the mass eigenvalue. As we shall see, it coincides with
the mass eigenvalue only at high temperatures if ${}^{}f_\varphi \ll f_a{}^{}$. 

Expanding potentials~\eqref{V_QCD} and~\eqref{V_mix} up to quadratic order in $\,a/f_a{}^{}{}^{}$ and 
${}^{}{}^{}\varphi/f_{\varphi}{}^{}$, we yield the mass mixing matrix,
\begin{eqnarray}\label{massmixing}
M_{\varphi a}^2 \,=\, 
m_{\varphi}^2 
\left(
\begin{array}{cc}
\,1 & f_{\varphi}/f_a \\[0.1cm]
\,f_\varphi/f_a & \,\,(f_\varphi/f_a)^2 \hspace{-0.04cm}
\end{array}
\right)
+
\left(
\begin{array}{cc}
\,0 & 0 \\[0.1cm]
\,0 & \,m_a^2(T) \hspace{-0.03cm}
\end{array}
\right)
~.
\end{eqnarray}
Upon diagonalizing $M_{\varphi a}^2{}^{}$, we obtain the heavy and light mass eigenstates $a_H$ and $a_L$ and their respective
masses $m_H(T)$ and $m_L(T)$\,:
\begin{eqnarray}\label{eigen}
a_H \,=\, \varphi \cos\xi + a \sin\xi ~, \quad
a_L  \,=\, -\,\varphi \sin\xi + a \cos\xi ~,
\end{eqnarray}
\vspace{-0.6cm}
\begin{eqnarray}
m_{H,L}^2(T)
\,=\, 
\frac{1}{2}  m_a^2(T) 
\scalebox{1.3}{\Bigg\{}
\hspace{-0.08cm}
1 + \mathcal{R}_m^2  \frac{m_a^2}{m_a^2(T)}
\scalebox{1.3}{\Bigg[}
1 + \mathcal{R}_f^2 
{}\pm
\sqrt{
\scalebox{1.1}{\bigg(}
1-\mathcal{R}_f^2 - \frac{1}{\mathcal{R}_m^2}\frac{m_a^2(T)}{m_a^2}
\scalebox{1.1}{\bigg)}^{\hspace{-0.15cm}2} 
+ 4\mathcal{R}_f^2} \,\,
\scalebox{1.3}{\Bigg]}
\hspace{-0.1cm}
\scalebox{1.3}{\Bigg\}}
\,,
\label{mass_eigenvalues}
\end{eqnarray}
where we have defined the following parameters,
\begin{eqnarray}
\aboxed{
~\mathcal{R}_f  \equiv\, 
\frac{f_{\varphi}}{f_a}~,\quad 
\mathcal{R}_m  \equiv\, 
\frac{m_{\varphi}}{m_a}}~
\end{eqnarray}
which parametrize the ALP decay constant and mass relative to the QCD axion.
The mixing angle $\xi = \xi(T)$ satisfies $0<\xi < \pi/2{}^{}$, and is given by
\begin{eqnarray}
\cos\xi \,=\, 
\sqrt{\frac{1}{2}
\scalebox{1.1}{\bigg[}
1+ \frac{\sin(2\xi)}{\tan(2\xi)} 
\scalebox{1.1}{\bigg]}}
\label{mixing_angle_1}
\end{eqnarray}
with
\begin{eqnarray}
\tan(2\xi) \,=\, 
\frac{2{}^{}\mathcal{R}_f}{1-\mathcal{R}_f^2 - \big(1/\mathcal{R}_m^2\big) \big[m_a^2(T)/m_a^2\big]}
~, \quad 
\sin(2\xi) \,=\,  \frac{2 {}^{} \mathcal{R}_f {}^{} m_{\varphi}^2}{m_H^2(T) - m_L^2(T)}
~.
\label{mixing_angle_2}
\end{eqnarray}

\begin{figure}[t!]
\centering
\hspace{-0.8cm}
\includegraphics[scale=0.6]{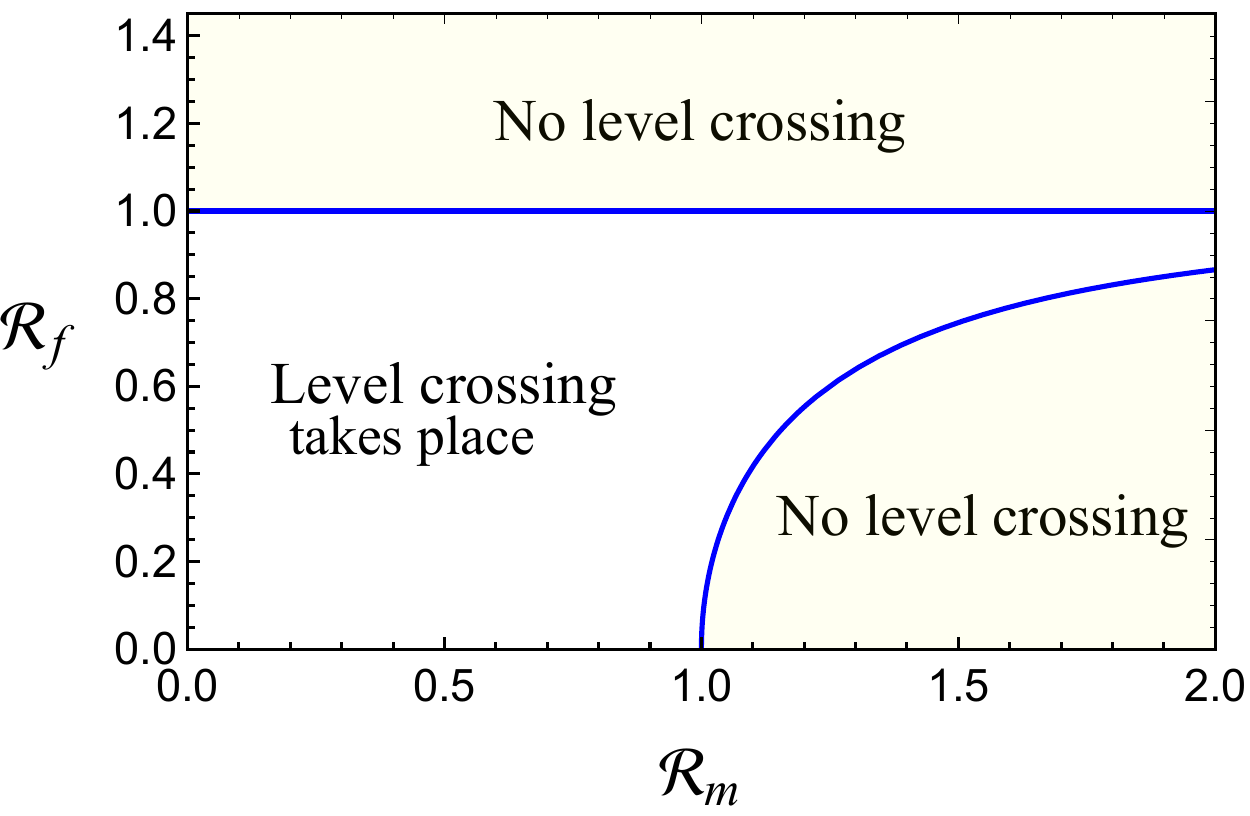}
\caption{
The level crossing takes place in the region between the two blue lines. See Eq.~(\ref{lc}).
}
\label{fig:lc}
\end{figure}

One interesting consequence of the mass mixing is that, as the cosmic temperature decreases,
the values of $m_H^2(T)$ and $m_L^2(T)$ approach to each other at high temperatures and 
move away from each other at low temperatures  under certain conditions. Let us call such behavior the level crossing. More formally, we define the level crossing as a situation in which the mass squared difference $m_H^2(T) - m_L^2(T)$ takes a minimum value for some finite value of ${}^{}T{}^{}$. From the condition, $d\big[m_H^2(T) - m_L^2(T)\big]/d{}^{}T = 0{}^{}$,
we find that the temperature at the level crossing $\,T_{\rm lc}$ is given by
\begin{eqnarray}
\frac{m_a^2\big(T_{\rm lc}\big)}{m_a^2} \,=\, 
\mathcal{R}_m^2
\scalebox{1.3}{\big(}
1-\mathcal{R}_f^2
\scalebox{1.3}{\big)}
~. \label{m_a_T_at_level_crossing}
\end{eqnarray}
Note that the above equation has a solution for $\,0 < m_a^2\big(T_{\rm lc}\big)/m_a^2 < 1$ if\,\footnote{
The definition of the level crossing is not unique. For example, we may define the level crossing as a situation in which the ratio of the mass eigenvalues is minimized~\cite{Daido:2015cba}, and one can derive a condition similar to Eq.~\eqref{lc_condition}.}
\begin{eqnarray}\label{lc_condition}
\mathcal{R}_f \,<\, 1 
\quad \text{and} \quad \mathcal{R}_m \,<\, \frac{1}{\sqrt{1-\mathcal{R}_f^2}}
~.
\label{lc}
\label{condition_level_crossing}
\end{eqnarray}
Therefore, the level crossing takes place only for the parameters satisfying Eq.~\eqref{condition_level_crossing}.
See Fig.~\ref{fig:lc}. 
In particular, if $\,\mathcal{R}_f \ll 1{}^{}$, the above conditions are reduced to $\,\mathcal{R}_m < 1\,$ and $\,\mathcal{R}_f \ll 1{}^{}$, i.e. both the ALP mass ${}^{}m_{\varphi}$ and its decay constant $f_{\varphi}$ should be smaller than those of the QCD axion.

%%%%%%%%%%%%%%%%%%%%%%%%%%%%%%%%%%%%%%%%%%%%%%%%%%
%%%%%%%%%%%%%%%%%%%%%%%%%%%%%%%%%%%%%%%%%%%%%%%%%%

To identify the peculiarity of the parameter region where the level crossing occurs,
let us see how the mass eigenstates behave asymptotically both at high and low temperatures.
For $m_a(T) \ll m_{\varphi}$ (i.e. at high temperatures), the mass eigenvalues from Eq.~\eqref{mass_eigenvalues} 
can be approximated as
\begin{eqnarray}
m_H^2\big(T\big) 
&\hspace{-0.15cm}\simeq\hspace{-0.15cm} &
\scalebox{1.3}{\big(}
1+\mathcal{R}_f^2
\scalebox{1.3}{\big)}m_{\varphi}^2 
\,\simeq\, 
\begin{cases}
\displaystyle
\,m_{\varphi}^2 & ~ \mathcal{R}_f \ll 1\\[0.2cm]
\displaystyle
\,\mathcal{R}_f^2  {}^{} m_{\varphi}^2 & ~\mathcal{R}_f \gg 1
\end{cases}
~,
\label{m_H_highT}
\\[0.5cm]
m_L^2(T) 
&\hspace{-0.15cm}\simeq\hspace{-0.15cm} &
\frac{m_a^2(T)}{1+\mathcal{R}_f^2} 
\,\simeq\, 
\begin{cases}
\displaystyle
\,m_a^2(T) & \mathcal{R}_f \ll 1
\\[0.2cm]
\displaystyle
\,\mathcal{R}_f^{-2}  m_a^2(T) & \mathcal{R}_f \gg 1 
\end{cases}
~.
\label{m_L_highT}
\end{eqnarray}
Note that, for $\mathcal{R}_f \gg 1{}^{}$, $m_H(T)$ becomes much larger than $m_{\varphi}{}^{}$, and $m_L(T)$ becomes much smaller than $m_a(T){}^{}$.
From Eqs.~\eqref{mixing_angle_1} and~\eqref{mixing_angle_2}, we can also obtain the asymptotic behavior of the mixing angle at high temperatures as
\begin{eqnarray}
\cos\xi (T) 
\,\simeq\, 
\frac{1}{\sqrt{1+\mathcal{R}_f^2}} 
\,\simeq\,
\begin{cases}
\displaystyle
\,1 
- \tfrac{1}{2}  \mathcal{R}_f^2
 & ~ \mathcal{R}_f \ll 1
\\[0.2cm]
\displaystyle
\,\mathcal{R}_f^{-1}  & ~ \mathcal{R}_f \gg 1
\end{cases}
~.
\label{cos_xi_highT}
\end{eqnarray}
On the other hand, the mass eigenvalues at zero temperature can be approximated as
\begin{eqnarray}
m_H^2 \big(T\to 0\big) 
&\hspace{-0.15cm}\simeq\hspace{-0.15cm} &
\begin{cases}
\displaystyle
\,m_a^2 
&~ \mathcal{R}_f \ll \mathcal{R}_m^{-1} \,\,\, \text{and} \,\,\, \mathcal{R}_m \ll 1  \\[0.25cm]
\displaystyle
\,m_{\varphi}^2 
&~\mathcal{R}_f \ll \mathcal{R}_m^{-1} \,\,\, \text{and} \,\,\,
\mathcal{R}_m \gg 1 \\[0.25cm]
\displaystyle
\,\scalebox{1.3}{\big(}
1+ \mathcal{R}_f^2 \scalebox{1.3}{\big)}
 m_{\varphi}^2 
&~\mathcal{R}_f \gg \mathcal{R}_m^{-1}
\end{cases}~,
\label{m_H_lowT}
\\[0.3cm]
m_L^2 \big(T\to 0\big) 
&\hspace{-0.15cm}\simeq\hspace{-0.15cm} &
\begin{cases}
\displaystyle
\,m_{\varphi}^2 
&~ \mathcal{R}_f \ll \mathcal{R}_m^{-1} \,\,\, \text{and} \,\,\, \mathcal{R}_m \ll 1 \\[0.25cm]
\displaystyle 
\,m_a^2 
&~ \mathcal{R}_f \ll \mathcal{R}_m^{-1} \,\,\, \text{and} \,\,\,  \mathcal{R}_m \gg 1 \\[0.2cm]
\displaystyle
\,\frac{m_a^2}{1+\mathcal{R}_f^2} 
&~ \mathcal{R}_f \gg \mathcal{R}_m^{-1}
\end{cases}
~.
\label{m_L_lowT}
\end{eqnarray}
Finally, the asymptotic behavior of the mixing angle at zero temperature is given by
\begin{eqnarray}
\cos\xi  \big(T\to 0\big) 
\,\simeq\,
\begin{cases}
\displaystyle
\,\mathcal{R}_m^2\mathcal{R}_f 
&~ \mathcal{R}_f \ll \mathcal{R}_m^{-1} \,\,\, \text{and} \,\,\, \mathcal{R}_m \ll 1 \\[0.2cm]
\displaystyle
\,\,1 - \tfrac{1}{2}  \mathcal{R}_f^2
&~ \mathcal{R}_f \ll \mathcal{R}_m^{-1} \,\,\, \text{and} \,\,\, \mathcal{R}_m \gg 1 \\[0.2cm]
\displaystyle
\,\frac{1}{\sqrt{1+\mathcal{R}_f^2}} 
&~ \mathcal{R}_f \gg \mathcal{R}_m^{-1}
\end{cases}
~.
\label{cos_xi_lowT}
\end{eqnarray}

\begin{figure}[t!]
\centering
\includegraphics[scale=0.26]{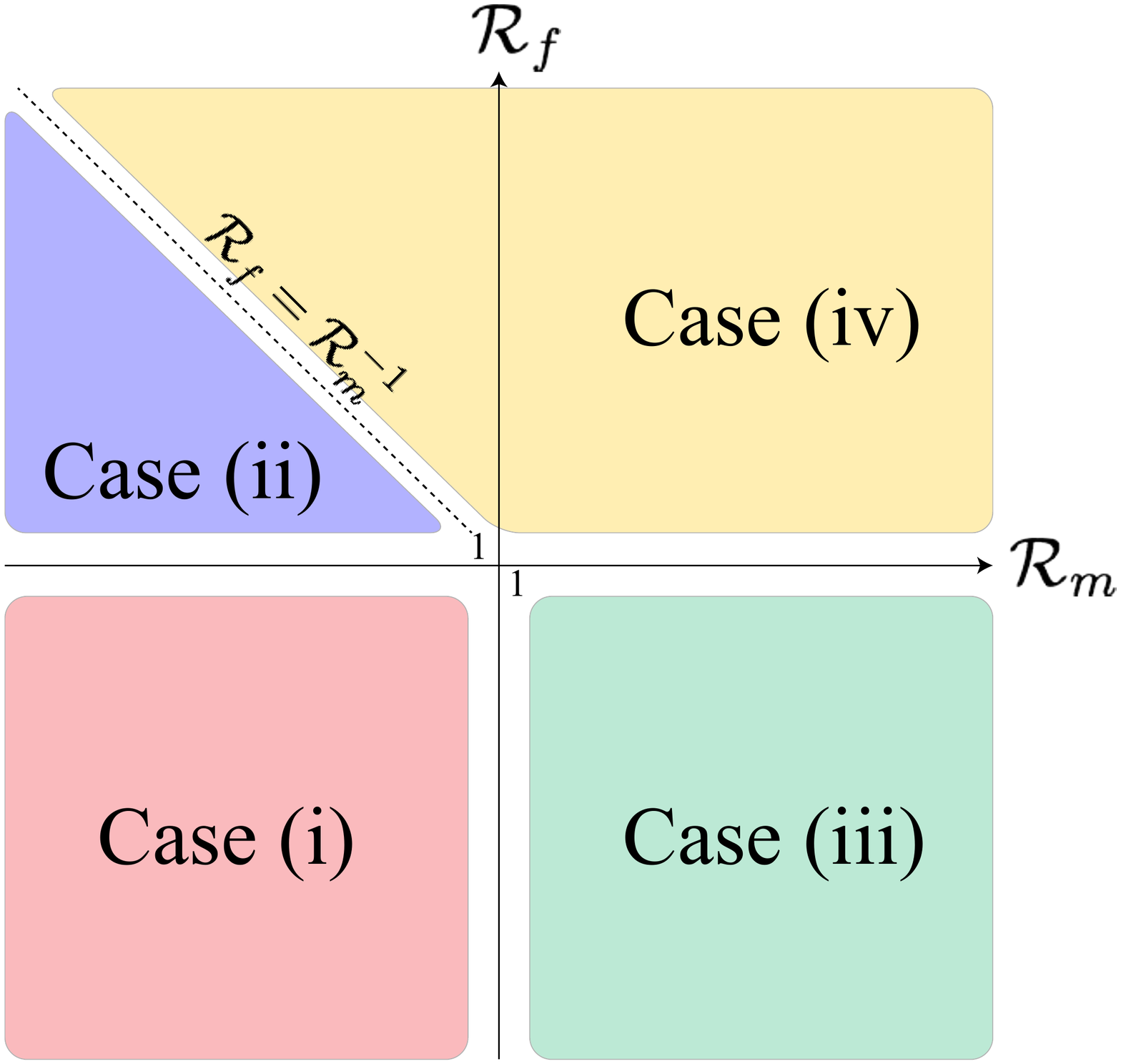}
\caption{
The regions corresponding to the cases (i)-(iv)  in the $\big({}^{}{\cal R}_m {}^{} , {\cal R}_f\big)$ plane
on a log-log scale. The level crossing occurs mostly in the case (i).
The cases (i) and (ii) are relevant for the discussion in Sec.~\ref{sec:cosmology}.
}
\label{fig:cases}
\end{figure}

According to Eqs.~\eqref{m_H_highT}-\eqref{cos_xi_lowT}, we can classify the behavior of the mass eigenvalues
into the four different cases as summarized in Table~\ref{tab:asymptotic_behavior}. The regions corresponding to the four cases are also shown in Fig.~\ref{fig:cases}. 
In the following sections, only the cases (i) and (ii) are relevant for our purpose since in cases (iii) and (iv) the heavier mass eigenvalue does not significantly change 
from high to low temperatures (see Table~\ref{tab:asymptotic_behavior}) and hence the heavy axion decouples at sufficiently low temperatures.
On the other hand, both mass eigenvalues evolve in a non-trivial manner in cases (i) and (ii). We will discuss more details in Sec.~\ref{sec:relic_DM}.
%%%%%%%%%%%%%%%%%%%%%%%%%%%%%%%%%%%%%%%%%%%%%%%%%%
\begin{table}[h]
\caption{Asymptotic behavior of the mass eigenstates and eigenvalues 
from high to low temperatures for the four different cases (i)-(iv).
}
\vspace{3mm}
\centering 
\begin{tabular}{| l |cccc|c|}
\hline
$\vphantom{|_|^|}$Cases & $a_H$ & $a_L$ & $m_H(T)$ & $m_L(T)$ \\[0.05cm]
\hline
\hline
$\vphantom{|_|^|}$(i) $\mathcal{R}_f \ll 1 \ll \mathcal{R}_m^{-1}$ & $\varphi \to a$ & $a \to -{}^{}\varphi$ & $m_{\varphi} \to m_a$ & $m_a(T) \to m_{\varphi}$ \\[0.1cm]
(ii) $1 \ll \mathcal{R}_f \ll \mathcal{R}_m^{-1}$ & $a \to a$ & $-{}^{}\varphi \to -{}^{}\varphi$ & $\mathcal{R}_f {}^{} m_{\varphi} \to m_a$ & $\mathcal{R}_f^{-1}m_a(T) \to m_{\varphi}$ \\[0.1cm]
(iii) $\mathcal{R}_f \ll 1\,$ and $\,\mathcal{R}_m \gg 1$ & $\varphi \to \varphi$ & $a \to a$ & $m_{\varphi} \to m_{\varphi}$ & $m_a(T) \to m_a$ \\[0.1cm]
(iv) $\mathcal{R}_f \gg 1\,$ and $\,\mathcal{R}_f \gg \mathcal{R}_m^{-1}$ & $a \to a$ & $-{}^{}\varphi \to -{}^{}\varphi$ & $\mathcal{R}_f {}^{} m_{\varphi} \to \mathcal{R}_f {}^{} m_{\varphi}$ & $\mathcal{R}_f^{-1}m_a(T) \to \mathcal{R}_f^{-1}m_a$ \\[0.12cm]
\hline
\end{tabular}
\label{tab:asymptotic_behavior}
\end{table}
%%%%%%%%%%%%%%%%%%%%%%%%%%%%%%%%%%%%%%%%%%%%%%%%%%

%%%%%%%%%%%%%%%%%%%%%%%%%%%%%%%%%%%%%%%%%%%%%%%%%%
%%%%%%%%%%%%%%%%%%%%%%%%%%%%%%%%%%%%%%%%%%%%%%%%%%

\begin{figure}[t!]
\centering
$\begin{array}{cc}
{
\includegraphics[scale=0.61]{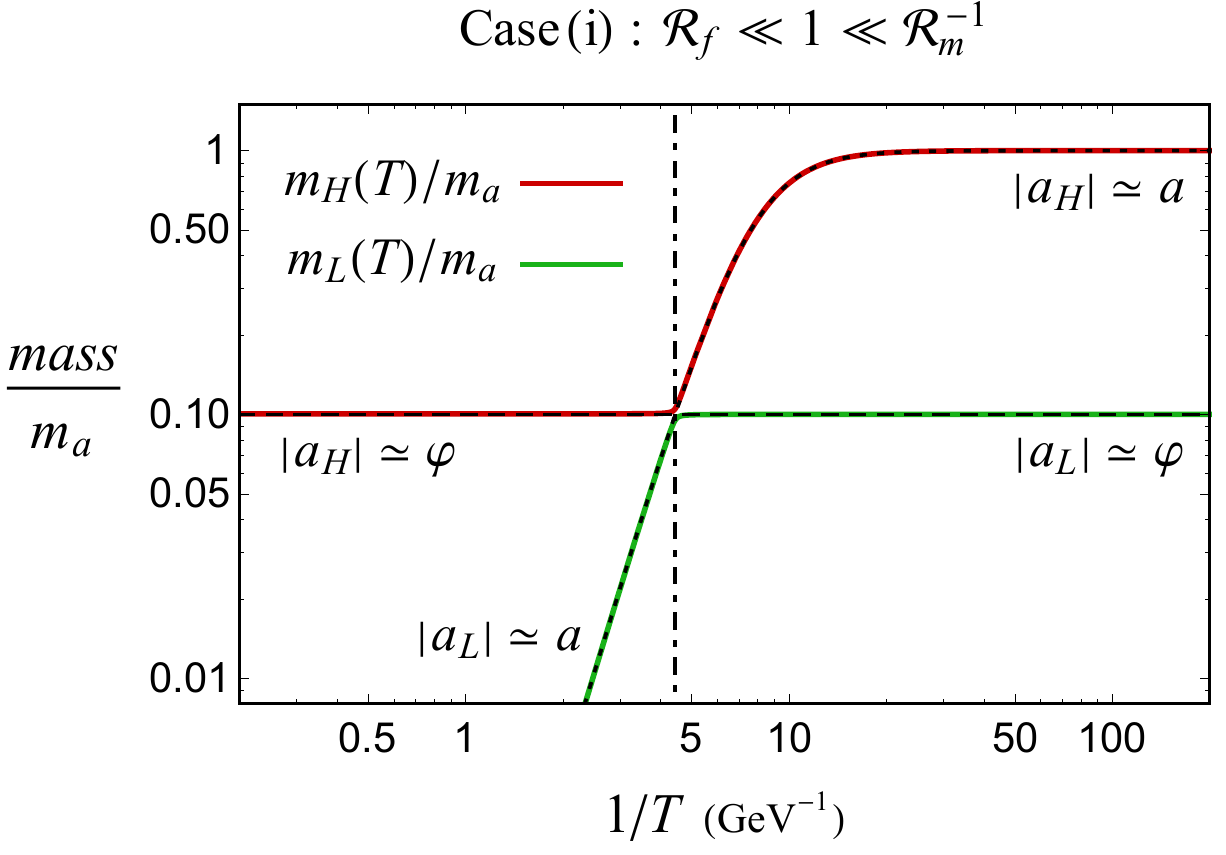}} \hspace{0.5cm}
{
\includegraphics[scale=0.61]{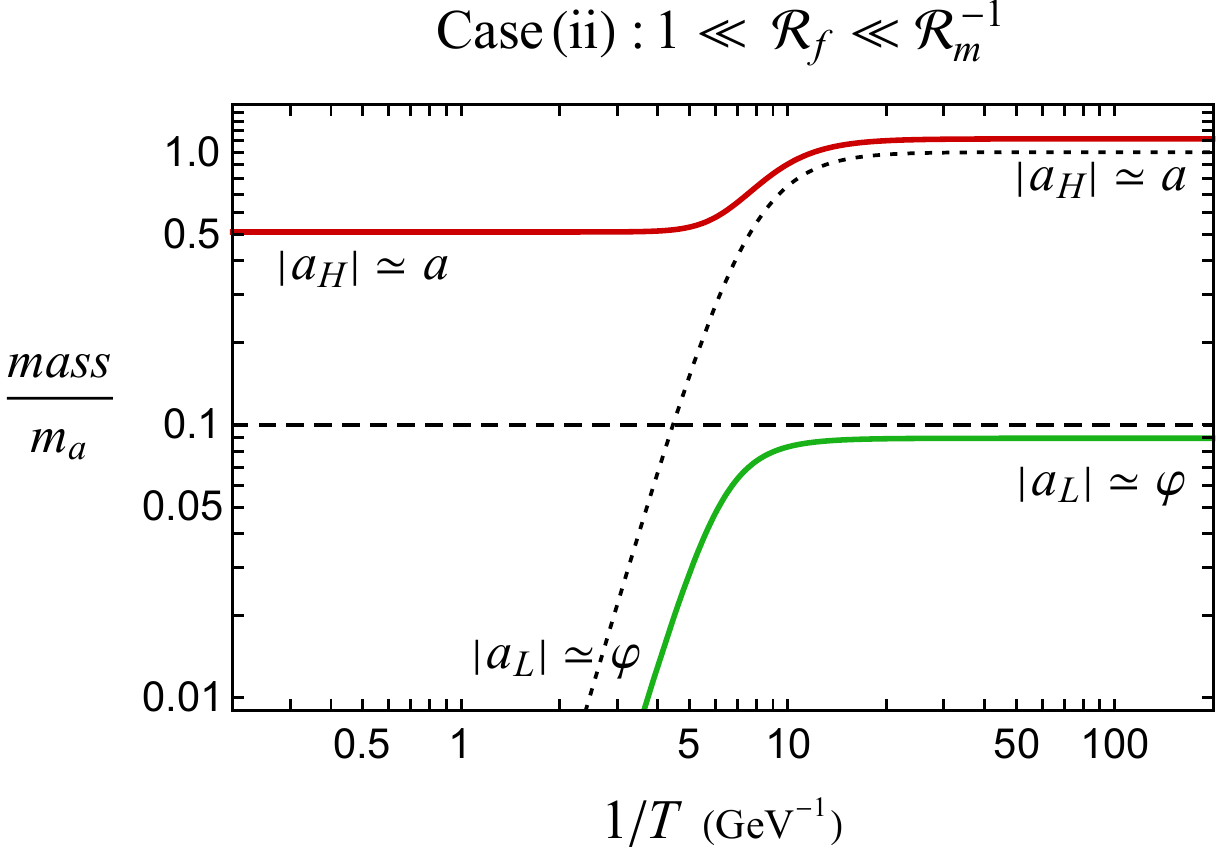}}\\[0.5cm]
{
\includegraphics[scale=0.61]{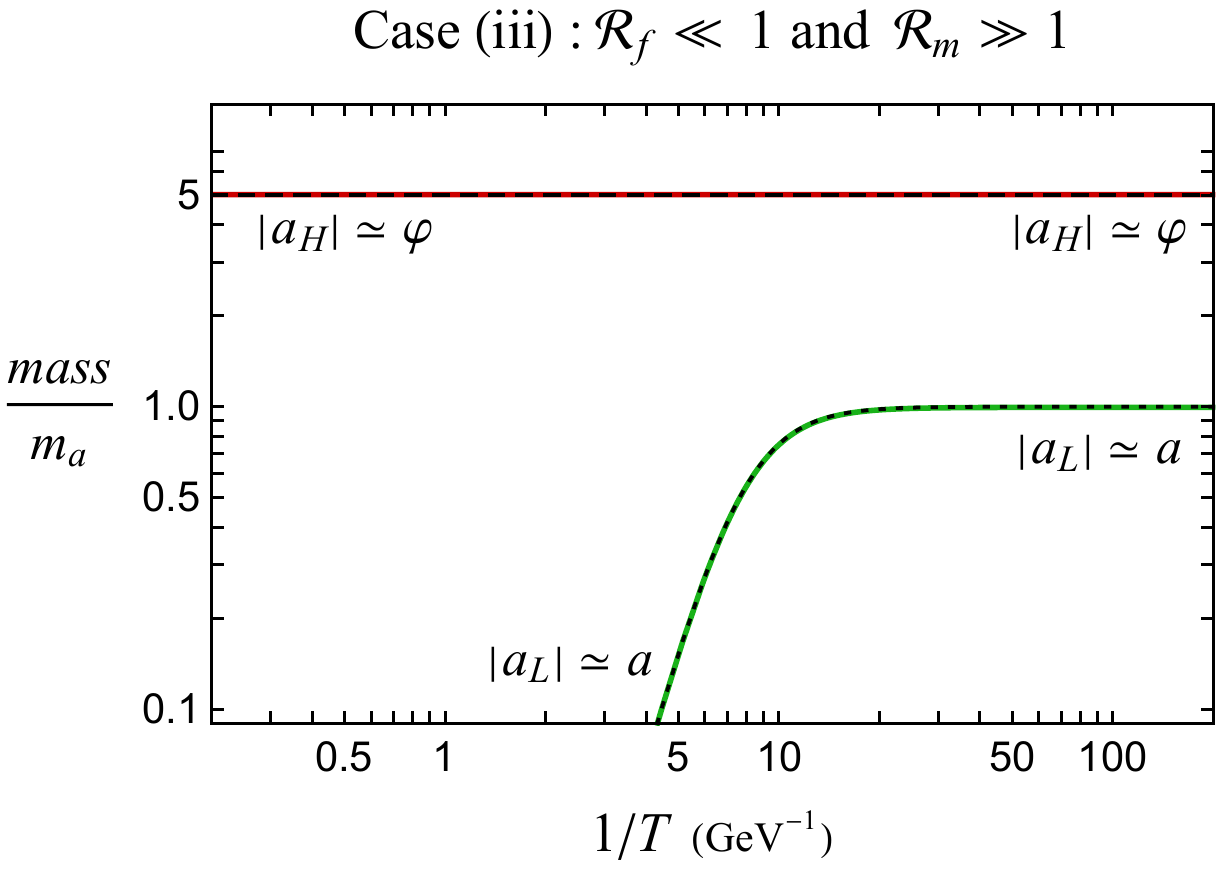}} \hspace{0.5cm}
{
\includegraphics[scale=0.61]{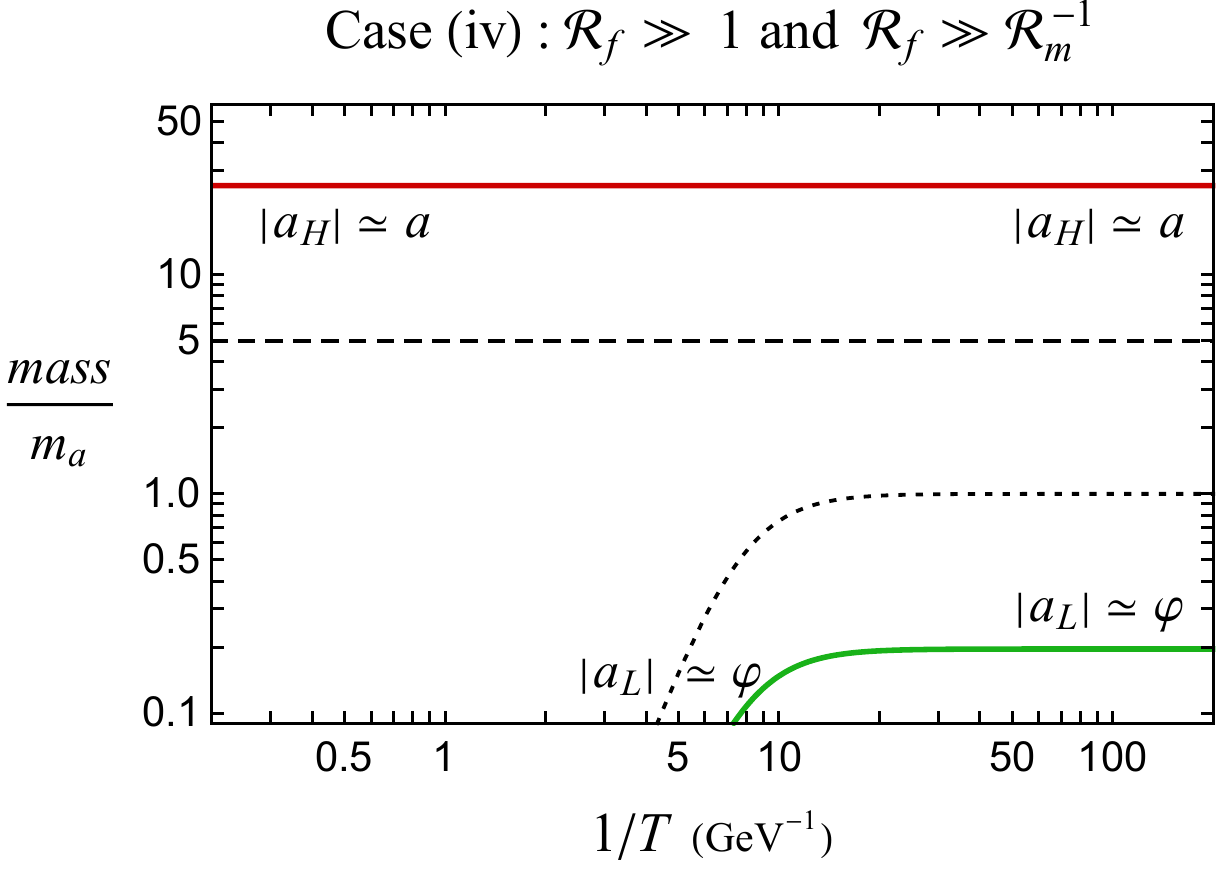}}
\end{array}$
\caption{Evolution of the mass eigenvalues as functions of temperature. 
In these panels, the red (green) solid line represents the heavier (lighter) mass eigenvalue, the dotted (dashed) black line denotes 
$m_a(T)\,(m_\varphi)$, and the dot-dashed black line indicates the temperature at the level crossing.
Here, all the masses are normalized by the zero temperature QCD axion mass, $m_a{}^{}$.
For each plot, we have set (i)~${\cal R}_m = 0.1{}^{}, {\cal R}_f = 0.1{}^{}$,
(ii)~${\cal R}_m = 0.1{}^{}, {\cal R}_f = 5{}^{}$, (iii)~${\cal R}_m = 5{}^{}, {\cal R}_f = 0.1{}^{}$, and
(iv)~${\cal R}_m = 5{}^{}, {\cal R}_f = 5{}^{}$,
which approximately correspond to the four different cases in Table~\ref{tab:asymptotic_behavior}.
(Mild hierarchies among the parameters are chosen for illustration purpose.)}
\label{fig:levelcrossing}
\end{figure}

\begin{figure}[h!]
\centering
\includegraphics[scale=0.61]{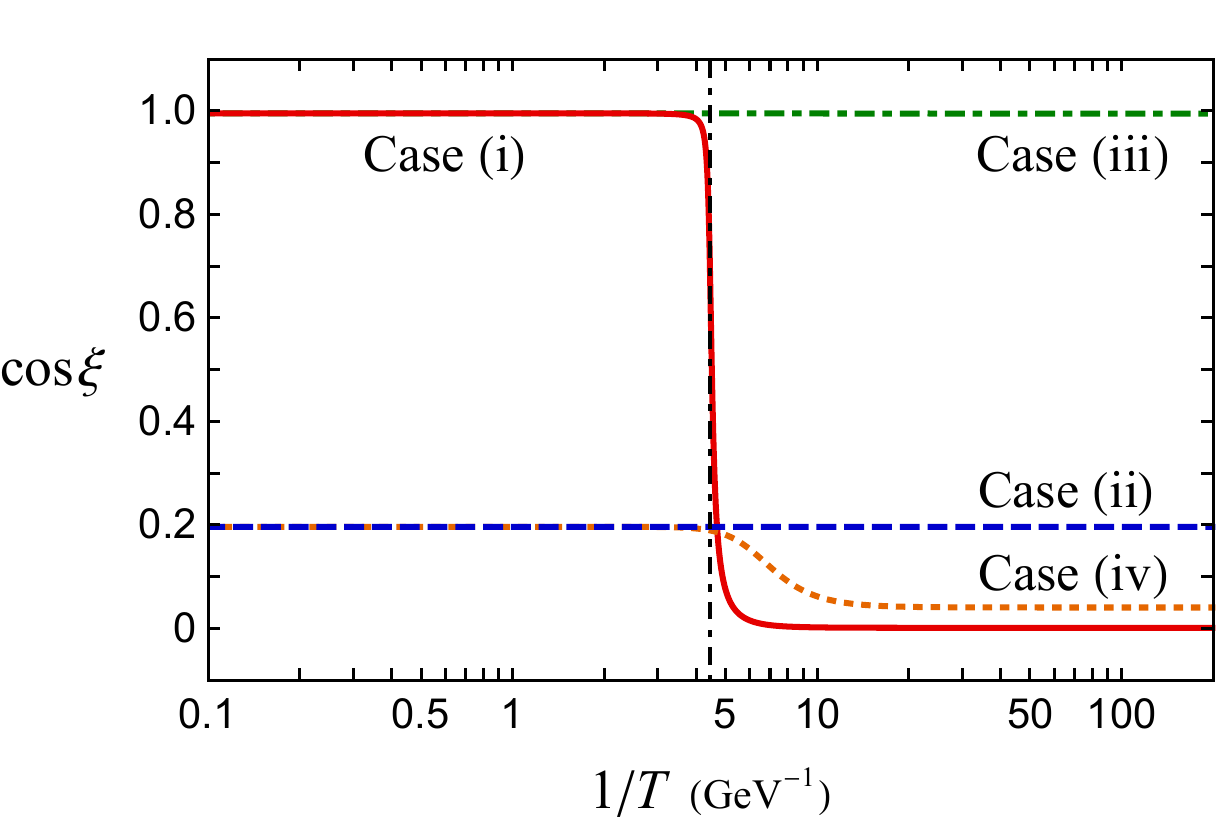}
\caption{Evolution of the mixing angle as functions of temperature
for the four different cases in Fig.~\ref{fig:levelcrossing}: solid for case (i), dashed for case (ii), dash-dotted for case (iii),
and dotted for case (iv). As shown in the figure, the sharp transition of the mixing angle only occurs in case (i),
which is contrasted with cases (ii)-(iv), where the mixing angle remains almost constant.}
\label{fig:levelcrossing-mixing}
\end{figure}

In Figs.~\ref{fig:levelcrossing} and \ref{fig:levelcrossing-mixing}, we show how the two mass eigenvalues and the mixing angle $\,\cos\xi\,$ evolve with the cosmic temperature.
Among these four cases, only the case (i) exhibits a peculiar behavior that the composition of the 
mass eigenstates drastically changes 
as they evolve from high to low temperatures.
Indeed, the level crossing necessarily takes places in 
the parameter region $\big(\mathcal{R}_f \ll 1 \ll \mathcal{R}_m^{-1}\big)$ for the case (i).
See Eq.~\eqref{condition_level_crossing}.
From Eqs.~\eqref{mixing_angle_1},~\eqref{mixing_angle_2} and~\eqref{m_a_T_at_level_crossing}, we see that the mixing becomes maximal at the level crossing,
\begin{eqnarray}\label{maximal_mixing}
\cos\xi\,\big|_{T\,=\,T_{\rm lc}} =\, \frac{1}{\sqrt{2}} ~,
\end{eqnarray}
while the mixing remains small otherwise.
Furthermore, the difference between the two mass eigenvalues at the level crossing reads
\begin{eqnarray}
m_H^2\big(T_{\rm lc}\big) - m_L^2\big(T_{\rm lc}\big)
\,=\,
2  {}^{} \mathcal{R}_f  {}^{} m_\varphi^2 ~,
\end{eqnarray}
which becomes much smaller than $m_a^2$ and $m_{\varphi}^2$ for $\mathcal{R}_f \ll 1$ and $\mathcal{R}_m \ll 1{}^{}$. 
As we will see in the next section, these properties are essential to realizing the adiabatic conversion between two mass eigenstates. 

We note that, in addition to the case (i), it is worth discussing the case (ii). This is because in this case the heavy (light) axion starts to oscillate in much earlier (later) time compared to the case (i), which also leads to a non-trivial effects on the final DM abundance. We will further elaborate on this point in Sec.~\ref{sec:relic_DM}.

%%%%%%%%%%%%%%%%%%%%%%%%%%%%%%%%%%%%%%%%%%%%%%%%%%

%%%%%%%%%%%%%%%%%%%%%%%%%%%%%%%%%%%%%%%%%%%%%%%%%%

%%%%%%%%%%%%%%%%%%%%%%%%%%%%%%%%%%%%%%%%%%%%%%%%%%
\section{Adiabatic conversion and cosmological axion abundances}
\label{sec:cosmology}
%%%%%%%%%%%%%%%%%%%%%%%%%%%%%%%%%%%%%%%%%%%%%%%%%%
In this section, we study the cosmological evolution of the QCD axion and ALP to evaluate their densities
and delineate the region in the parameter space where the sum of their 
contributions explain 
the observed DM abundance.
In doing so, we clarify the conditions for the adiabatic conversion to take place,
which plays an important role to determine their abundances. 

%%%%%%%%%%%%%%%%%%%%%%%%%%%%%%%%%%%%%%%%%%%%%%%%%%
\subsection{Adiabatic conversion of the QCD axion and ALP}
\label{sec:angle_density}
\setcounter{equation}{0}
%%%%%%%%%%%%%%%%%%%%%%%%%%%%%%%%%%%%%%%%%%%%%%%%%%
In the previous study~\cite{Kitajima:2014xla}, the adiabatic condition between the QCD axion and ALP was simply described 
without checking its validity in the whole relevant parameter space including the case in which
there is a hierarchy between two decay constants $f_a$ and $f_{\varphi}$.
In this subsection, we will generalize it 
based on an argument on adiabatic invariants for multiple harmonic oscillators~\cite{nHO} 
and confirm it by using the numerical analysis. To this end, let us start with
the equations of motion for $a$ and  $\varphi$, which are given by \cite{Kitajima:2014xla}
\begin{align}\label{fieldEOM1}
&\ddot{a} +
3   H  \dot{a} +
m_a^2(T) f_a  \sin\hspace{-0.05cm}\bigg(\frac{a}{f_a}\bigg) +
\frac{m_\varphi^2  f_\varphi^2}{f_a}
\sin\hspace{-0.05cm}
\bigg(\frac{a}{f_a} + \frac{\varphi}{f_\varphi}\bigg)
\,=\,0 ~,\\[0.15cm]
\label{fieldEOM2}
&\ddot{\varphi} +
3   H  \dot{\varphi} +
m_\varphi^2  f_\varphi
\sin\hspace{-0.05cm}
\bigg(\frac{a}{f_a} + \frac{\varphi}{f_\varphi}\bigg)
\,=\,0~.
\end{align}
Here and in what follows we focus on the evolution of the homogenous QCD axion and ALP fields
in the radiation-dominated universe. 
For convenience, we introduce the effective angles, $\theta$ and $\Theta$, defined by
\begin{eqnarray}\label{angles}
\theta \,=\, \frac{a}{f_a} ~,\quad
\Theta \,=\, \frac{a}{f_a} + \frac{\varphi}{f_\varphi} ~.
\end{eqnarray}
In terms of the effective angles, the equations of motion read
\begin{align}\label{stheta}
& \ddot{\theta} +
3 H  \dot{\theta} +
m_a^2(T) \sin\theta +
m_\varphi^2  {\cal R}^2_f
\sin \Theta
\,=\,0~,\\[0.15cm]
\label{btheta}
& \ddot{\Theta} +
3   H\dot{\Theta} +
m_a^2(T) \sin\theta +
m_\varphi^2\Big(1+{\cal R}^2_f\Big)
\sin\hspace{-0.03cm}
\Theta
\,=\,0 ~,
\end{align}
which we will numerically solve under the initial conditions, $\theta(t_0) = \theta_0{}^{}, \Theta(t_0) = 
\Theta_0{}^{}$, and $\dot{\theta}(t_0) = \dot{\Theta}(t_0) = 0{}^{}$, where $t_0$ is some initial time well before
the QCD axion or ALP starts to oscillate.

After the light and heavy axions start to oscillate, i.e., $H  <  m_L(T){}^{}, m_H(T){}^{}$, 
the potential can be well approximated by the quadratic one since
their oscillation amplitudes decrease with the cosmic expansion.\footnote{
This is not always the case. If the axion starts to oscillate at around the level crossing, 
the axion may go over the potential hill at the level crossing 
and continue to run until it is trapped by one of the potential minima, 
{\it the axion roulette}~\cite{Daido:2015bva,Daido:2015cba}.
}
Let us define the comoving axion numbers of the heavy and light mass eigenstates, 
$N_H$ and $N_L$ as
\begin{eqnarray}\label{conumber}
N_H(T) \,\equiv\, \frac{\rho_H(T)}{m_H(T)  s(T)}  ~,\quad
N_L(T) \,\equiv\, \frac{\rho_L(T)}{m_L(T)   s(T)} ~,
\end{eqnarray}
where $s(T) = 2\pi^2  g_s(T) {}^{}{}^{} T^3/45$ is the entropy density 
with ${}^{}g_s(T)$ being the effective entropic degrees of freedom, and $\rho_{H,L}(T)$ is the energy density of the heavy or light
mass eigenstate. 
If there were not for the mass mixing, then both ${}^{}N_H$ and $N_L$  would be adiabatic invariants and so they would be
separately conserved. 
In the presence of the mixing, however, the situation is more involved since there are multiple time scales over which 
the potential changes. In addition to the mass eigenvalues, the mixing angle is also time-dependent, and in particular,
it exhibits a very sharp transition at the level crossing (see Fig.~\ref{fig:levelcrossing-mixing}).

The time evolution of the adiabatic invariants for multiple harmonic oscillators was studied in Ref.~\cite{nHO}.
Let  $t_{\rm ext}$ denote the typical time scale over which the oscillation frequencies change significantly.
The authors of Ref.~\cite{nHO} studied the time evolution of the quanta number of each oscillation 
eigenmode (like $N_H$ and $N_L$) using the Glauber variables.
It was shown there that the total quanta number (which corresponds to $N_H+N_L$ in our case) 
is conserved if $\,t_{\rm ext}$ is much longer than the oscillation period of each eigenmode. 
Even in this case, each quanta number (i.e. $\hspace{-0.1cm}N_H$ or $N_L$ in our case) is not necessarily conserved,
if $\,t_{\rm ext}$ is shorter than or comparable to the time scale determined by the difference of the angular
frequencies, i.e., the beat frequency of the two harmonic oscillators.\footnote{
The Glauber variables are the analog of the annihilation and creation operators, $(\alpha, \alpha^*)$.
They are written in terms of the canonical variables of the harmonic oscillator(s), $(q,p)$, as 
$\alpha \sim \sqrt{\omega} {}^{} q + i p/\sqrt{\omega}$, 
where $\omega$ is the angular frequency. For a constant $\omega$, $\alpha$ evolves like $\alpha \sim e^{-i \omega t}$.
For multiple harmonic oscillators, the quanta number of the $n$-th eigenmode
is given by $\alpha_n^* \alpha_n$. When the angular frequencies depend on time,  $d {}^{}\alpha_n/d {}^{}t$ involves 
terms including $\alpha_m$ and $\alpha_m^*$ with $m \ne n$. Therefore, the time evolution of the quanta number
of the $n$-th eigenmode, 
$d {}^{}  |\alpha_n|^2/d{}^{}t$, naturally involves differences between the angular frequencies, $\omega_n - \omega_m$, for $m \ne n$.
} 

Now, let $\,t_{\rm ext}{}^{}$ be the time scale over which the potential changes significantly. If there are multiple time scales, $t_{\rm ext}$ 
is chosen to be the shortest one among them. For example, in the case where the level 
crossing takes place, 
it is of order the Hubble time for most of the time, but it is given by the time interval for the mixing angle to 
change at the level crossing.
In any case, $t_{\rm ext}$ is always proportional to the Hubble time, as the cosmic expansion is the only source of the time-dependence. 
Applying the above argument 
to the present case, the adiabatic condition for both $N_H$ and $N_L$ to be conserved reads
\begin{eqnarray}
t_{\rm ext} \,\gg\, \mathrm{max}\Bigg[\frac{2\pi}{m_L(T)}, \frac{2\pi}{m_H(T) - m_L(T)}\Bigg] ~.
\label{adiabatic_condition}
\end{eqnarray}

The adiabatic conversion takes place if both $N_H$ and $N_L$ are separately conserved at the level
crossing. At the level crossing, $t_{\rm ext}$ is given by
\begin{eqnarray}
t_{\rm ext} \,=\, \left|\frac{1}{\cos\xi(T)}\frac{d\cos\xi(T)}{d{}^{}t}\right|^{-1}_{{}^{}T \,=\, T_\text{lc}} ~.
\label{t_ext_level_crossing}
\end{eqnarray}
The second term in the right-handed side of Eq.~(\ref{adiabatic_condition}) gives the dominant contribution 
in this case, for $\mathcal{R}_f \ll 1{}^{}$,  it reads
\begin{eqnarray}
\frac{2\pi}{m_H(T_\text{lc}) - m_L(T_\text{lc})} \,\simeq\, \frac{2\pi}{\mathcal{R}_f  {}^{} m_\varphi} ~.
\end{eqnarray}
Using Eqs.~\eqref{axionmass} and \eqref{mixing_angle_1}, 
and $\mathcal{R}_f \ll 1{}^{}$, we derive the adiabatic condition at the level crossing 
as
\begin{eqnarray}
\label{adiabatic_condition_gamma}
\boxed{~\gamma \,\equiv\,
\beta {}^{}{}^{} \mathcal{R}_f \, \sqrt{\frac{m_\varphi}{H(T_\text{lc})}} 
\,\gg 1 \,}
\end{eqnarray}
where we define
\begin{eqnarray}
\beta \,\equiv\,  \sqrt{\frac{1}{n\pi} 
\scalebox{1.1}{\bigg[}
1+\frac{1}{3}\frac{d\ln g_s(T)}{d\ln T}
\scalebox{1.1}{\bigg]}
}\,\Bigg|_{{}^{}T \,=\, T_\text{lc}} ~.
\end{eqnarray}

In general, once the condition~\eqref{adiabatic_condition} is satisfied, 
we expect that both $N_L$ and $N_H$ are conserved after both the heavy and light axions start to oscillate.
This is not the case if the adiabatic condition is violated. 
Note that here we implicitly assume that $t_{\rm ext}$ is larger than the first term $2\pi/m_L(T)$ [and hence $2\pi/m_H(T)$]
in the right-handed side of Eq.~\eqref{adiabatic_condition}
since otherwise one cannot define the numbers $N_L$ and $N_H$ unambiguously.
Then, we can consider the case where $t_{\rm ext}$ becomes smaller than $2\pi/[m_H(T)-m_L(T)]$ but still larger than $2\pi/m_L(T)$.
In this case, while the total quanta number $N_H+N_L$ is conserved, $N_H-N_L$ is not necessarily conserved.
Typically this can happen at the level crossing where $m_H^2 - m_L^2$ takes the minimal value and $t_{\rm ext}$ given by Eq.~\eqref{t_ext_level_crossing}
becomes sufficiently small.
In other words, once the condition~\eqref{adiabatic_condition_gamma} is violated,
$N_H-N_L$ is no longer conserved at the level crossing.

In the previous work~\cite{Kitajima:2014xla}, 
the adiabatic condition was considered to be $\,m_\varphi/H(T_\text{lc}) \gg 1{}^{}$. 
However, the refined condition~\eqref{adiabatic_condition_gamma} contains a factor of $\,\mathcal{R}_f$
apart from the numerical coefficient $\beta{}^{}$. This is because we have included the beat frequency in addition to the
angular frequency of each mass eigenstate in the adiabatic condition. This implies that, even if 
$m_\varphi/H(T_\text{lc}) \hspace{-0.06cm} \gg \hspace{-0.06cm} 1{}^{}$,
the adiabatic condition could be broken if $\,\mathcal{R}_f$ is sufficiently small. 

Note that the adiabatic condition~\eqref{adiabatic_condition_gamma} makes sense only for the parameters satisfying 
Eq.~\eqref{condition_level_crossing} (or the case (i) in the previous section), 
since the level crossing occurs only in this case.
In the other cases (ii), (iii), and (iv) (see Fig.~\ref{fig:cases} and Table~\ref{tab:asymptotic_behavior}), 
the original condition~\eqref{adiabatic_condition} is trivially satisfied after the light and heavy axions start to oscillate.
Since the level crossing does not take place in these cases, their time evolution is simple and nothing more than just the
two oscillating scalar fields. 

\begin{figure}[t!]
\begin{center}
\includegraphics[scale=0.6]{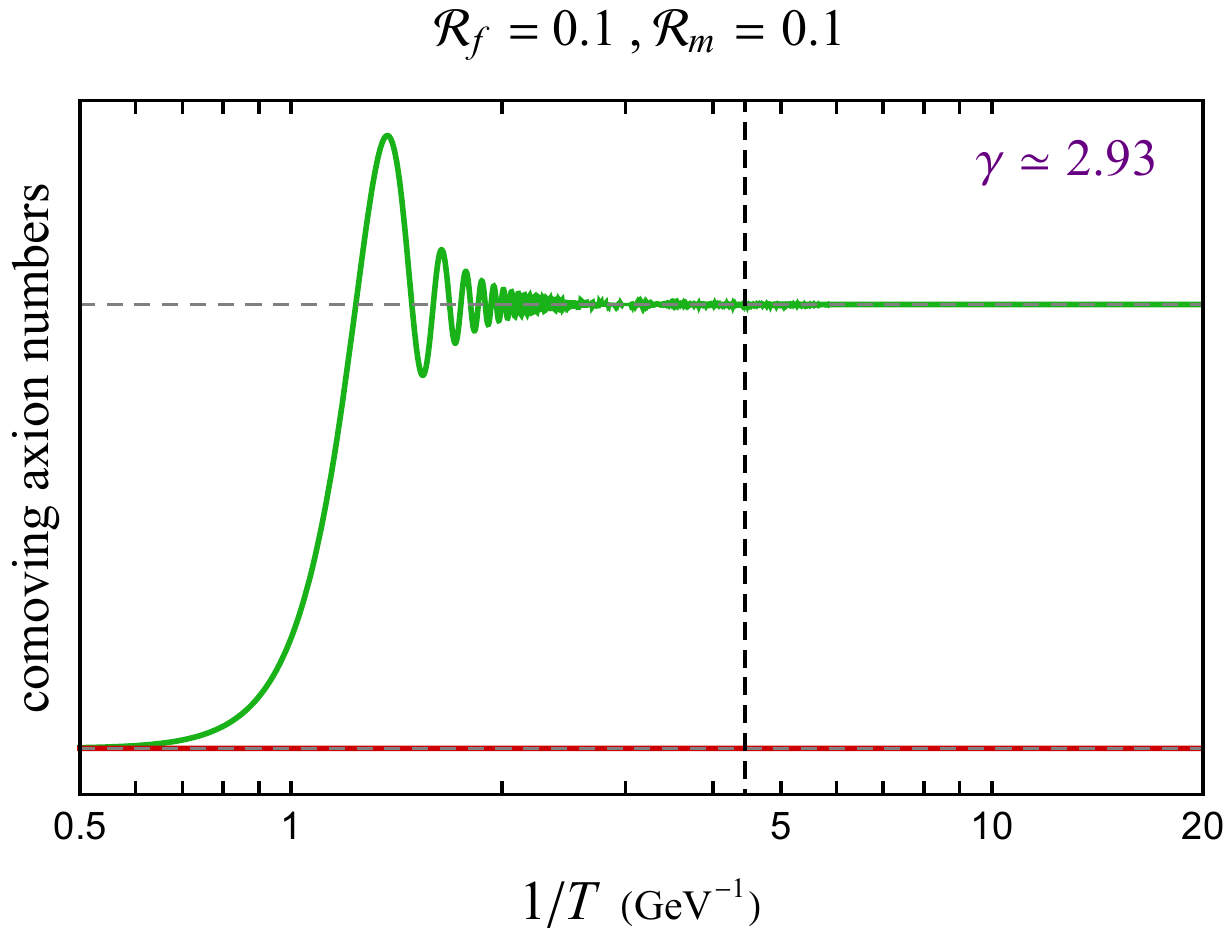}
\hspace{0.3cm}
\includegraphics[scale=0.6]{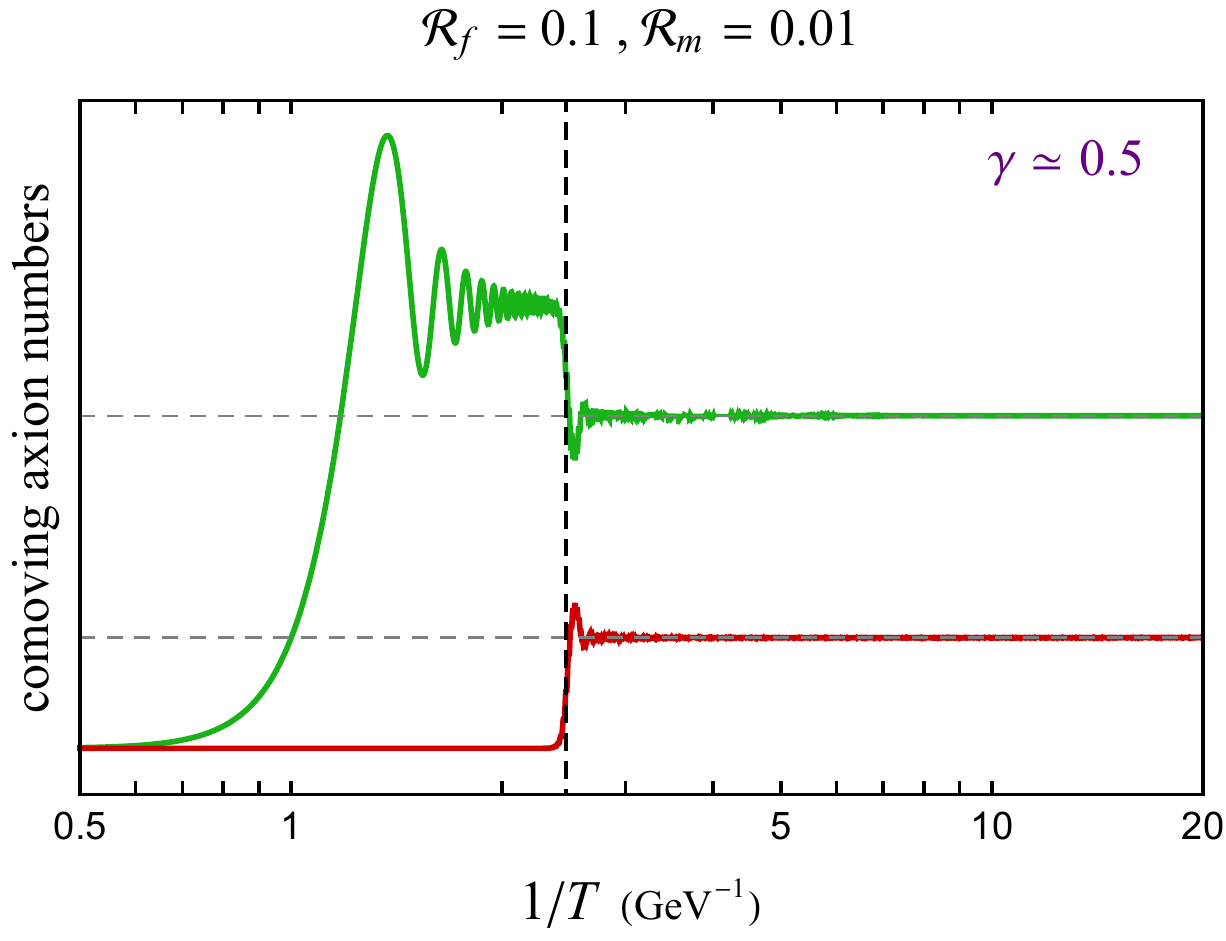}
\\[0.3cm]
\includegraphics[scale=0.6]{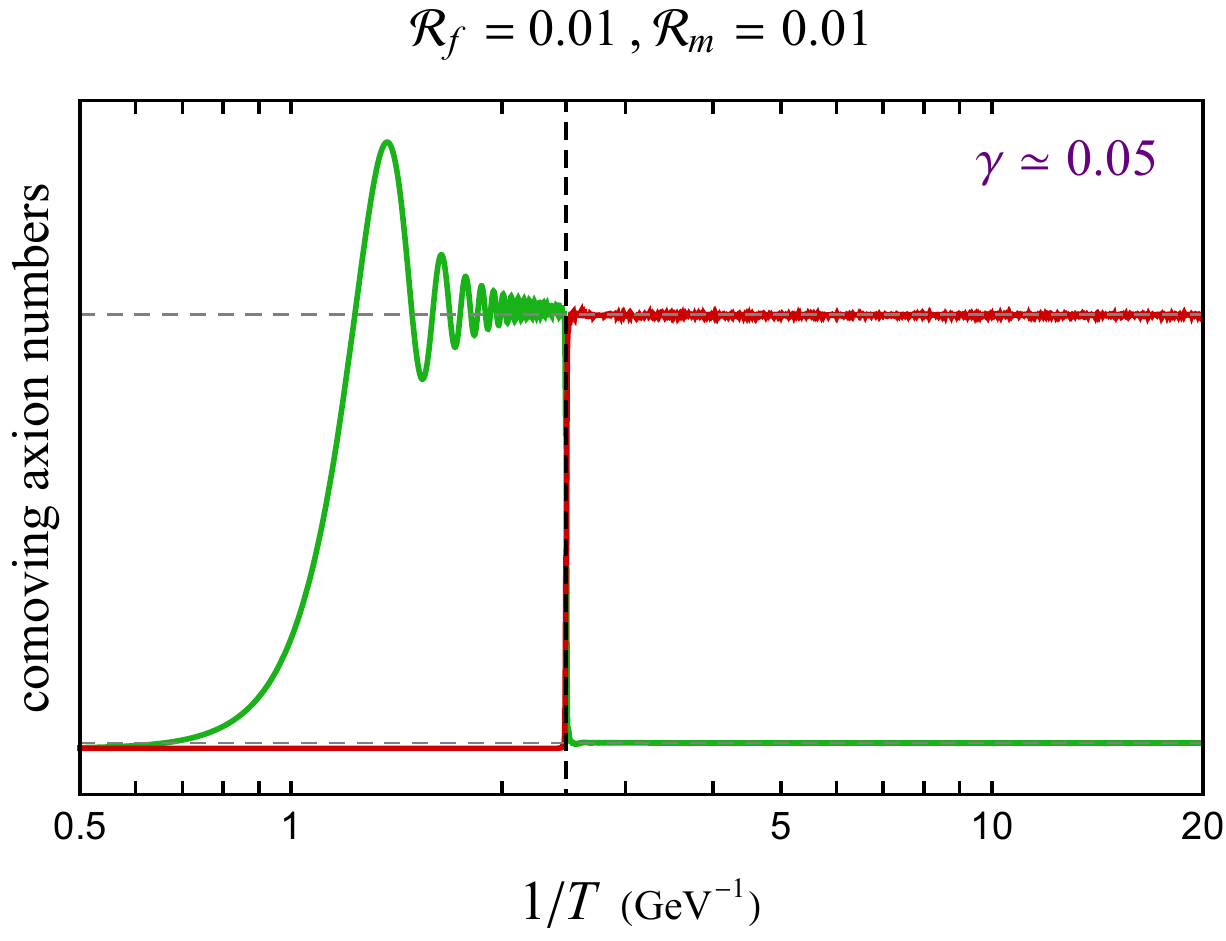}
\caption{Evolution of the comoving axion numbers as functions of temperature for various values of the 
parameters satisfying Eq.~\eqref{condition_level_crossing}.
The values of $\gamma$ evaluated from Eq.~\eqref{adiabatic_condition_gamma} is also shown.
The red (green) solid line represents $N_H \,(N_L)$, and the vertical dashed black line is the level crossing point
defined by Eq.~(\ref{m_a_T_at_level_crossing}).
Here we have fixed $\theta_0 = 1{}^{}$, $\Theta_0 = 0{}^{}$, and $f_a = 10^{12} \,\text{GeV}$.
}
\label{fig:number}
\end{center}
\vspace{-0.3cm}
\end{figure}

In Fig.~\ref{fig:number}, we show our numerical results of the comoving axion numbers as functions of the temperature for
several choices of the parameters satisfying Eq.~\eqref{condition_level_crossing}.
In these plots, we omit the scale in the vertical axes since the overall normalization of the comoving axion number is arbitrary. 
As expected, the comoving axion numbers are not conserved at the level crossing when ${}^{}\gamma {}^{} \lesssim{}^{} 1{}^{}$. 
(Note that they are not conserved in the second panel even if ${}^{}m_\varphi/H(T_\text{lc}) \sim 10 \gg 1$.) The adiabatic conversion takes place in the first panel.
In spite of the rapid change of the mixing angle at the level crossing, the comoving axion numbers are conserved.
In this case, the quantity $N_L\,$($N_H$) initially identified as the number of the QCD axion (ALP) is converted into 
that of the ALP (QCD axion) until the present time [see case (i) in Table~\ref{tab:asymptotic_behavior}],
which affects their cosmological abundance significantly as we shall see next.

%%%%%%%%%%%%%%%%%%%%%%%%%%%%%%%%%%%%%%%%%%%%%%%%%%
%%%%%%%%%%%%%%%%%%%%%%%%%%%%%%%%%%%%%%%%%%%%%%%%%%

%%%%%%%%%%%%%%%%%%%%%%%%%%%%%%%%%%%%%%%%%%%%%%%%%%
\subsection{Dark matter abundance}
\label{sec:relic_DM}
%%%%%%%%%%%%%%%%%%%%%%%%%%%%%%%%%%%%%%%%%%%%%%%%%%
Now let us evaluate the DM abundance. In our model both the light and heavy axions could contribute to DM, and the total DM abundance is given by the sum of their contributions,
\begin{eqnarray}
\Omega_\text{DM}  h^2 
\,=\,\Omega_H  h^2+  \Omega_L  h^2  ~,\quad
\end{eqnarray}
\vspace{-0.8cm}
\begin{eqnarray}\label{relicLH}
\Omega_H h^2 \,=\,
\frac{m_H s_0}{\rho_{c,0}}N_H ~,\quad
\Omega_L h^2 \,=\,
\frac{m_L s_0}{\rho_{c,0}}N_L ~, 
\end{eqnarray}
where $m_{H,L}  \equiv  m_{H,L}(T\to 0){}^{}$,  and $\rho_{c,0} \, (s_0)$ is the present critical (entropy) density. 
Note that $N_H$ and $N_L$ in Eq.~\eqref{relicLH} must be evaluated well after the level crossing since 
the adiabatic condition may be broken during the level crossing. 

The adiabatic conversion has a significant impact on the DM abundance. Let us first consider the case (i) 
with $\,\mathcal{R}_f \ll 1{}^{}$ and ${}^{}\mathcal{R}_m \ll 1{}^{}$, as the level crossing mostly takes place in this case. 
We focus on the evolution of the light mass eigenstate, $a_L{}^{}$.
What is peculiar is that ${}^{}a_L$ behaves like the QCD axion before the level crossing, but it behaves like
the ALP after the level crossing (see Table~\ref{tab:asymptotic_behavior}).  In particular, its mass is equal to ${}^{}{}^{}m_\varphi$ in the present universe. 
If the adiabatic conversion takes place at the level crossing, the comoving axion number ${}^{}N_L$ is
conserved. Therefore, the contribution of ${}^{}a_L{}^{}$ to the DM abundance is smaller by the mass ratio, ${\cal R}_m 
= m_\varphi/m_a {}^{}$, than without the mass mixing. This is because the QCD axion with the similar comoving number 
would obtain the mass $\,m_a$  if there were not for the mixing. 

\begin{figure}[t!]
\centering
$\begin{array}{cc}
{
\includegraphics[scale=0.6]{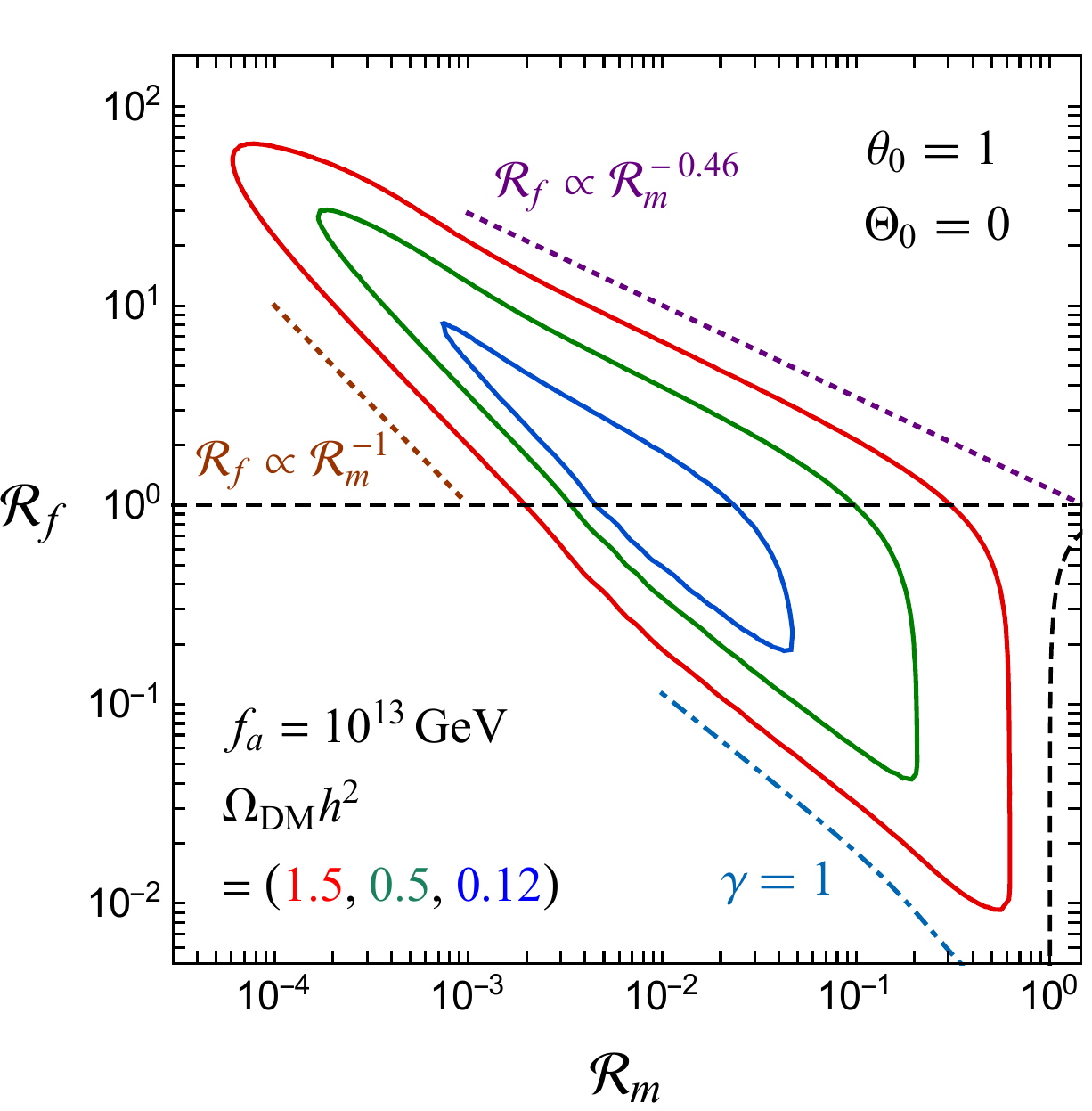}}
\hspace{0.8cm}
{
\includegraphics[scale=0.6]{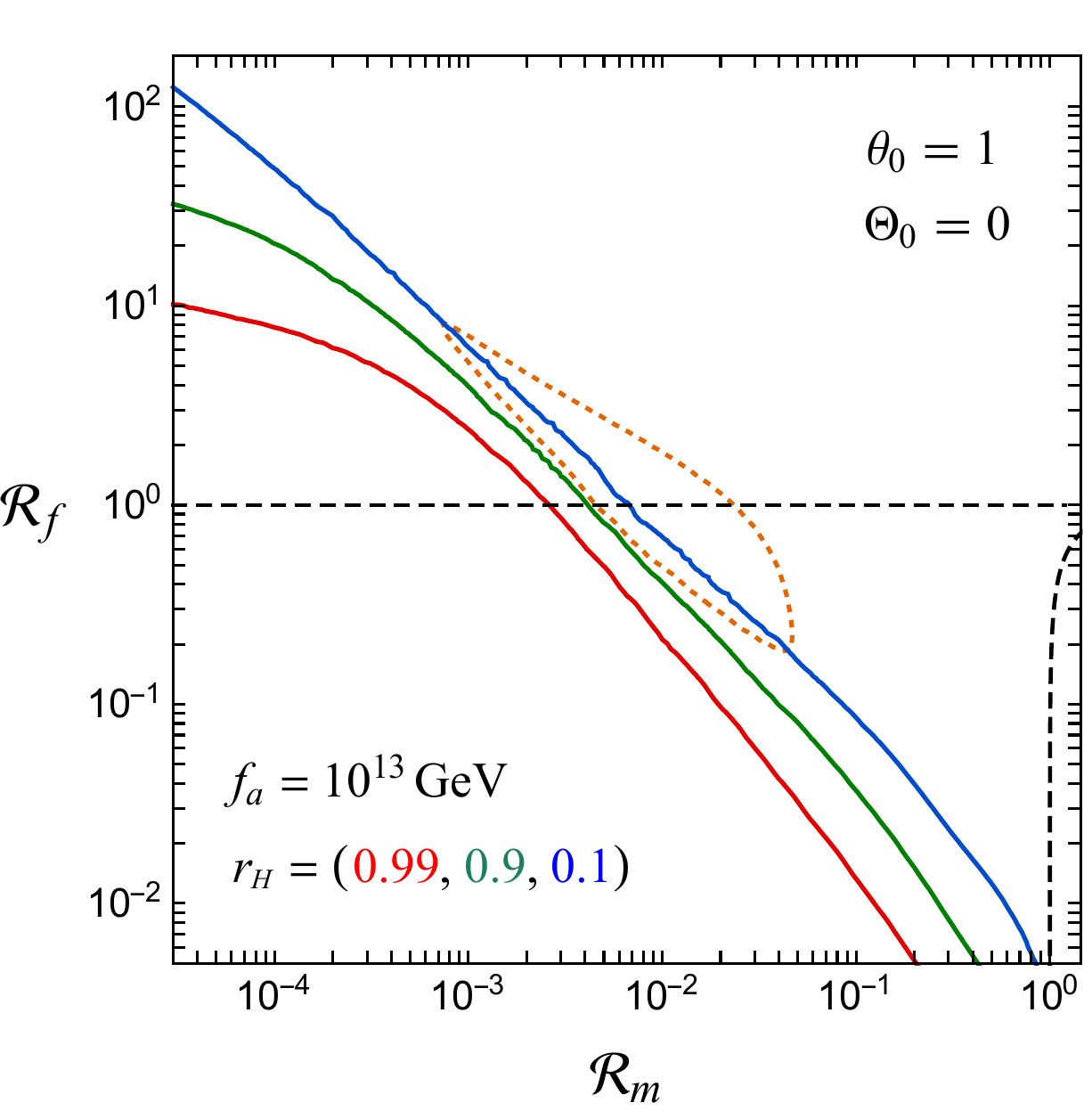}}
\end{array}$
\caption{Contours of the DM abundance $\Omega_{\rm DM}h^2 = 1.5{}^{}, 0.5{}^{}$, and $0.12$ (left) from
outermost to innermost and the relative fraction of the heavy axion $r_H = 0.99{}^{}, 0.9{}^{}$, and $0.1$ (right)
from left to right. 
We set $f_a = 10^{13}$\,GeV and adopt the initial condition ${}^{}\theta_0 = 1$ and $\Theta_0 = 0{}^{}$.
In both panels, we also show the condition (\ref{lc})
for the level crossing as the black dashed line. In the left panel, we show the adiabatic condition
$\gamma = 1{}^{}$, ${\cal R}_f \propto {\cal R}_m^{-0.46}$, and ${\cal R}_f \propto {\cal R}_m^{-1}$, which explain
the shape of the contours (see the text). In the right panel, the orange dashed contour represents
$\,\Omega_{\rm DM} h^2 = 0.12{}^{}$. 
The area outside the blue contour (left) and dashed orange contour (right) are ruled out by the observational value of DM abundance.
}
\label{fig:RfvsRm}
\end{figure}

We have numerically followed the evolution of the QCD axion and ALP for a broad range of $\,{\cal R}_f$
and ${\cal R}_m$ and
calculated their abundances.\footnote{
It is $m_\varphi$ and $f_\varphi$ that are actually varied in the plane of $\big({\cal R}_m{}^{}, {\cal R}_f\big)$ as we fix $f_a$ in the following.
} In Fig.~\ref{fig:RfvsRm}, we show the contours of the DM abundance, 
$\Omega_\text{DM}  h^2 = 0.12{}^{}, 0.5{}^{}$, and 1.5 in the left panel and  the contours of the relative fraction of the
heavy eigenstates,
\begin{eqnarray}
r_H \,=\, \frac{\Omega_H}{\Omega_L + \Omega_H} ~,
\end{eqnarray}
in the right panel. 
To see if the QCD axion abundance is indeed suppressed by the mass ratio,
we adopt $f_a = 10^{13}$\,GeV for which in the absence of the mass mixing
the QCD axion generated by the realignment mechanism would give a too large contribution to DM 
unless the initial angle is smaller than unity.
In other words, there would be no allowed parameter region
if there were not for the mass mixing. We also adopt the initial condition ${}^{}\theta_0 = 1$ and $\Theta_0 = 0{}^{}$
for which $a_H$ is already at the potential minimum at sufficiently high temperatures.\footnote{
Precisely speaking, this is the case only if the heavy mass eigenvalue becomes equal to the Hubble
parameter before the temperature-dependent axion mass ${}^{}m_a(T)$ becomes relevant. 
} 
This enables us to focus on the evolution of $a_L{}^{}$. We shall see later how the results are modified
for different initial conditions.

%%%%%%%%%%%%%%%%%%%%%%%%%%%%%%%%%%%%%%%%%%%%%%%%%%
%%%%%%%%%%%%%%%%%%%%%%%%%%%%%%%%%%%%%%%%%%%%%%%%%%

As one can see from the left panel of Fig.~\ref{fig:RfvsRm}, there is a triangle-shaped region where one can explain DM. The region outside the blue line is excluded since the total axion abundance exceeds the observed DM abundance ($\Omega_\text{DM} h^2 \simeq 0.12$). In fact, the DM is dominated by $a_L$ except for some region near ${}^{}{\cal R}_f = 1$ along the left boundary 
(see the right panel of Fig.~\ref{fig:RfvsRm}).
Such viable region appears (partly) due to the adiabatic conversion.
We explain below how the boundary of the region is determined. 

First, we notice that the viable region is fully
contained in the region $\mathcal{R}_m < 1$ and $\mathcal{R}_f < \mathcal{R}_m^{-1}$, which 
roughly correspond to the cases (i) and (ii). We can split the region into $\mathcal{R}_f < 1$ (case (i))
and $\mathcal{R}_f >1$ (case (ii)).
Let us begin with the region with $\mathcal{R}_f < 1{}^{}$. In order to suppress the axion abundance by the
adiabatic conversion, the mass ratio $\,\mathcal{R}_m$ needs to be less than unity.
If we decrease the value of $\,\mathcal{R}_m$ from unity for fixed $\mathcal{R}_f \,(<1){}^{}$, 
at some point the suppression becomes sufficient to explain the observed DM abundance.
This corresponds to the right vertical contours shown in the left panel of Fig.~\ref{fig:RfvsRm}.
Then, if we further decrease the value of $\mathcal{R}_m{}^{}$, the abundance becomes more suppressed,
but at a certain point, the adiabatic condition becomes violated, since $\,\gamma\,$ defined in Eq.~(\ref{adiabatic_condition_gamma}) decreases as $\,m_\varphi$ decreases for fixed ${\cal R}_f{}^{}$.
(Note that $H(T_{\rm lc})$ increases as $\,m_\varphi$ decreases.) 
Once the adiabatic condition is broken, $N_L$ is no longer conserved, and its large fraction is
converted to $N_H$. For the adopted parameters, the production of the heavy mode leads to the
DM overproduction, which explains the bottom-left part of the contours.
Indeed, the bottom-left boundary runs in parallel with the line of $\gamma = 1{}^{}$, 
below which the adiabatic condition is violated. The production of the heavy mode in the region ${}^{}\gamma < 1$
can also be seen in the right panel of Fig.~\ref{fig:RfvsRm}.

Next, let us consider the region with ${}^{}\mathcal{R}_f > 1{}^{}$.
In this case,  the level crossing does not take place and the adiabatic condition is trivial. In other words,
$N_H$ and $N_L$ are separately conserved after both axions start to oscillate. Therefore, it is important to
know when and how much those heavy and light axions are produced. 

We start with the upper-right region of the triangle-shaped boundary.
The heavy mass eigenvalue is given by ${}^{}{\cal R}_f {}^{} m_\varphi$ at high temperatures,
and it grows as ${}^{}m_a(T)$ at low temperatures. They satisfy ${\cal R}_f {}^{} m_\varphi <m_a$ 
in the region we consider.  Near the upper-right boundary,
the heavy axion mass  becomes equal to the Hubble parameter well before the 
temperature-dependent axion mass, $m_a(T){}^{}$, turns on.
For the present choice of the initial condition, the heavy axion already sits at the potential minimum 
(in the limit of $m_a(T) \to 0$), and so, the heavy axion has a negligible abundance. This can also be seen in
the right panel of Fig.~\ref{fig:RfvsRm}. 
So let us focus on the light axion whose mass evolves as ${\cal R}_f^{-1} m_a(T)$ at high temperatures,
and asymptotes to $\,m_\varphi$ at low temperatures. 
When the light axion mass  becomes 
comparable to the Hubble parameter, the light axion starts to oscillate. Let us denote as ${}^{}H_{\rm osc}$ 
the Hubble parameter at the onset of the oscillation. Afterwards, $N_L \propto H_{\rm osc}^{-1/2} \hspace{-0.05cm}f_\varphi^{{}^{}2}\,$
is conserved. Therefore, the DM abundance is
proportional to $m_\varphi {}^{} H_{\rm osc}^{-1/2} \hspace{-0.05cm} f_\varphi^{{}^{}2} {}^{}\propto m_\varphi f_\varphi^{{}^{}(2n+5)/(n+2)}$,
where $n = 4.08$ is given in Eq.~(\ref{axionmass}). This explains the dependence of the 
upper-right boundary, ${\cal R}_f \propto {\cal R}_m^{-0.46}$. 

Finally, let us consider what happens if one decreases $\,{\cal R}_m$ from the upper-right boundary for a fixed ${\cal R}_f{}^{}$. In this case,  the heavy axion mass becomes lighter, and at a certain point,
it becomes equal to the Hubble parameter after the QCD axion mass turns on.
So it starts to oscillate when $m_a(T) \sim H$, and it oscillates along the QCD axion $a{}^{}$. 
For $f_a = 10^{13}$\,GeV, thus produced QCD axion exceeds the observed DM abundance for the
oscillation amplitude of order unity, and therefore,
there will be no allowed region once the heavy axion starts to oscillate when $m_a(T)\hspace{-0.05cm} \sim \hspace{-0.05cm} H$. 
The critical point is where $\,{\cal R}_f {}^{} m_\varphi \sim m_a(T) \sim H$. Since the middle and right-handed side
of this relation are independent of $f_\varphi$ or $m_\varphi{}^{}$, the upper-left boundary satisfies ${\cal R}_f \hspace{-0.06cm} \propto \hspace{-0.06cm}
{\cal R}_m^{-1}$. Around the critical point, the heavy axion is slightly produced by the realignment mechanism
explained above. Since the boundary is determined by the heavy axion abundance,
it sensitively depends on the initial condition in contrast to the other boundaries. We will see below how this boundary is modified for different initial conditions.

In the right panel of Fig.~\ref{fig:RfvsRm}, we show the contours of $\,r_H {}^{}={}^{} 0.99{}^{},0.9{}^{}$, and ${}^{}0.1{}^{}$, 
as well as the contour of $\,\Omega_{\rm DM} = 0.12$ (orange dashed line). 
The region outside the orange dashed line is excluded since the total axion abundance
exceeds the observed DM abundance.
If one looks at the left-boundary of the dashed line, one can see that the fraction of the heavy axion increases from 
${}^{}0.1{}^{}$ to ${}^{}0.9{}^{}$ as $\,{\cal R}_f$ increases to unity, and then decreases from ${}^{}0.9{}^{}$ to ${}^{}0.1{}^{}$ as $\,{\cal R}_f$ further increases from unity. This behavior may be understood as follows.\footnote{
While our estimate on the axion abundance could be modified near $\,{\cal R}_f = 1$
as this is the boundary of the cases (i) and (ii),  the interpretation given here should hold when ${\cal R}_f$
much larger or smaller than unity.
} For  $\,{\cal R}_f \lesssim 1{}^{}$, the light axion abundance becomes more suppressed as $\,{\cal R}_m$ decreases, and so, the heavy axion needs to compensate it to explain
the observed DM abundance. This explains why $r_H$ increases along the orange dashed line as ${}^{}{\cal R}_f$
increases (or ${}^{}{\cal R}_m$ decreases). 
On the other hand, for  ${\cal R}_f \gtrsim 1{}^{}$, the light axion abundance is proportional to $m_\varphi f_\varphi^{{}^{}(2n+5)/(n+2)}$, which increases along the orange dashed line which scales as ${\cal R}_f
\propto {\cal R}_m^{-1}$. Therefore, the heavy axion fraction decreases along the orange dashed line as ${}^{}{\cal R}_f$
increases (or $\,{\cal R}_m$ decreases). The heavy axion can be produced either by the weak violation of the adiabatic condition or by a small contribution of the realignment mechanism.

\begin{figure}[h]
\begin{center}
\hspace{-1cm}
\includegraphics[scale=0.6]{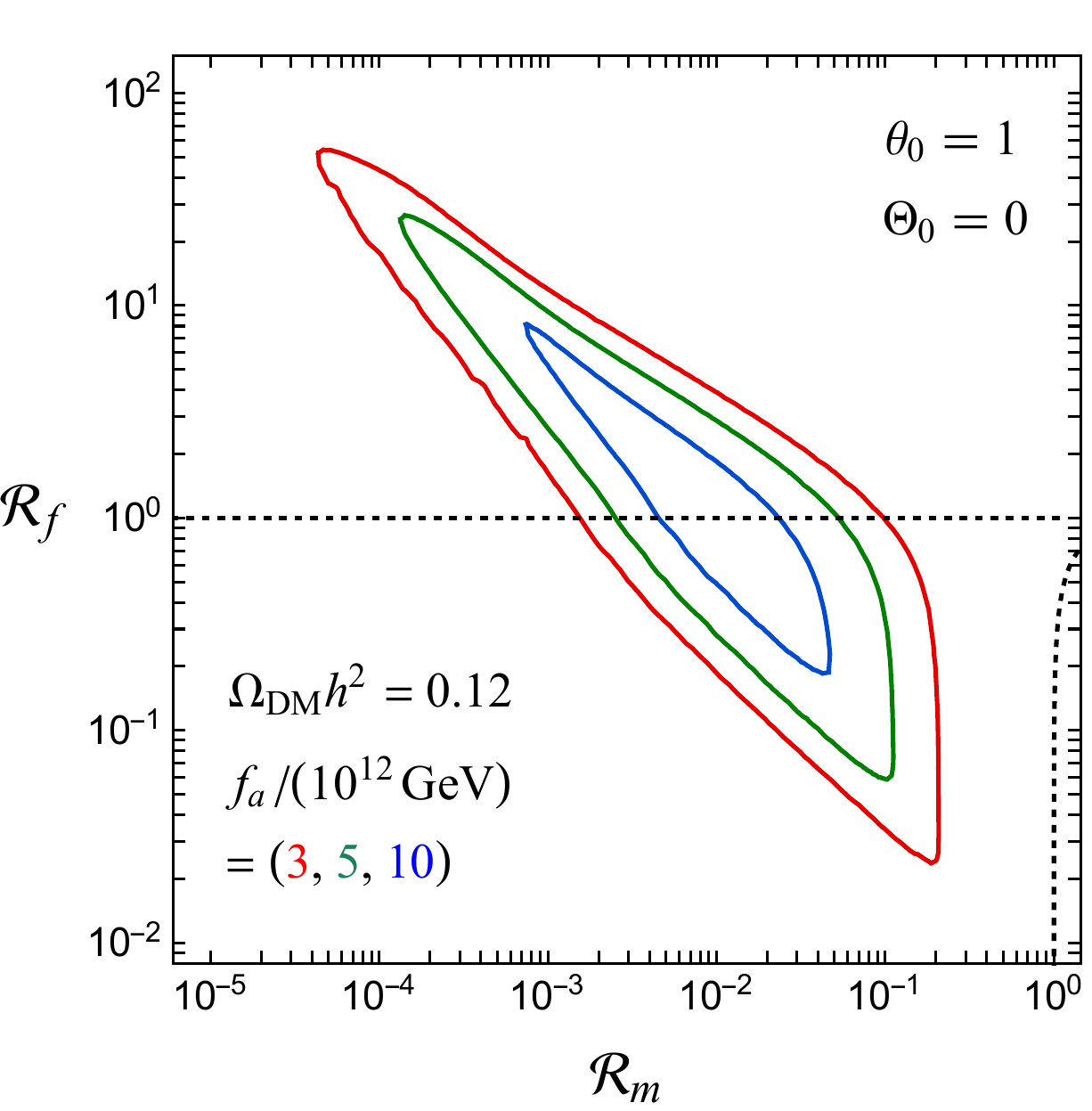}
\caption{Contours of the DM abundance $\,\Omega_{\rm DM}h^2 = 0.12\,$  for 
$f_a = (3{}^{}, 5{}^{}, 10)\times 10^{12}$\,GeV and the initial conditions ${}^{}\theta_0 = 1$ and ${}^{}\Theta_0 = 0$. }
\label{fig:RfvsRm_rH}
\end{center}
\end{figure}

In Fig.~\ref{fig:RfvsRm_rH} we show the contours of $\,\Omega_{\rm DM}h^2 = 0.12\,$ for different values of
${}^{}f_a = (3{}^{},5{}^{}, 10) \times 10^{12}$\,GeV. The region expands as $f_a$ decreases 
since the required suppression of the DM abundance becomes smaller.  

%%%%%%%%%%%%%%%%%%%%%%%%%%%%%%%%%%%%%%%%%%%%%%%%%%
%%%%%%%%%%%%%%%%%%%%%%%%%%%%%%%%%%%%%%%%%%%%%%%%%%

\begin{figure}[h]
\begin{center}
\hspace{-1cm}
\includegraphics[scale=0.6]{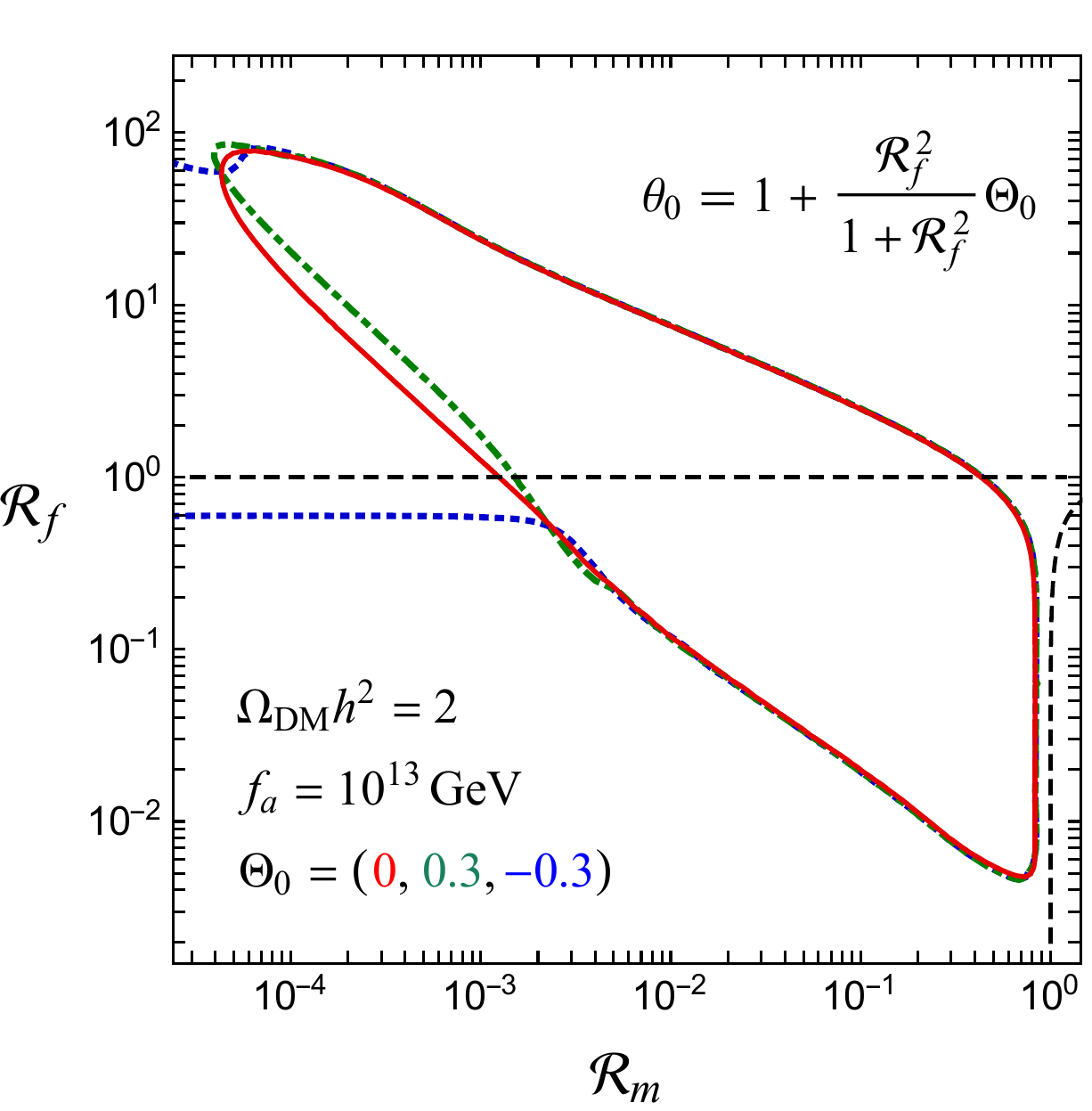}
\caption{Contours of the DM abundance $\,\Omega_{\rm DM}h^2 \,=\,2\,$  for 
different initial angles, $\Theta_0 = -\,0.3{}^{}, 0{}^{}$, and $0.3{}^{}$. We also vary ${}^{}\theta_0{}^{}$ 
as $\,\theta_0 = 1+\big[{\cal R}_f^2/\big(1+{\cal R}_f^2\big)\big]\Theta_0{}^{}$, and adopt $f_a = 10^{13}$\,GeV.
 The choice of $\Omega_{\rm DM}h^2 ={}^{} 2{}^{}$ is for illustration
purpose. The shape of the contours of $\,\Omega_{\rm DM}h^2 {}^{}={}^{} 0.12{}^{}$ for the above initial angles and decay constant looks very similar, implying that our results are robust. 
}
\label{fig:Rf_Rm_Theta}
\end{center}
\end{figure}

We show the contour of ${}^{}{}^{}\Omega_{\rm DM} h^2 = 2$ for different initial angles in Fig.~\ref{fig:Rf_Rm_Theta}.
To ease the comparison we choose such initial angles that only the heavy axion is varied, while
the light one along the minimum of  the potential (\ref{eq:mixV}) remains the same.
This is realized by varying $\Theta_0$ with $\,\theta_0 = 1 + \big[{\cal R}_f^2/\big(1+{\cal R}_f^2\big)\big]\Theta_0{}^{}$.
As one can see from Fig.~\ref{fig:Rf_Rm_Theta}, the boundaries are almost the same for the three cases
where they are determined by the light axion abundance.  The exception is the left boundary with ${\cal R}_f > 1$
where it is determined by the heavy axion abundance. For $\Theta_0 = 0.3{}^{}$, the heavy axion abundance increases
relative to $\Theta_0 = 0{}^{}$, since its effective amplitude is determined by $\theta_0{}^{}$. For $\Theta_0 = -\,0.3{}^{}$, the heavy axion abundance decreases relative to $\Theta_0 = 0{}^{}$, and the total abundance falls short of 
$\,\Omega_{\rm DM}h^2 = 2{}^{}$. This explains the behavior of the blue dotted line in Fig.~\ref{fig:Rf_Rm_Theta}. 
In the region between the blue dotted lines, the DM abundance is approximately constant,
and it is given by $\,\Omega_{\rm DM}h^2 \simeq 1.2{}^{}$. This is because the DM is dominated by the heavy axion, i.e. the QCD axion whose abundance is independent of $\,m_\varphi$ and $f_\varphi{}^{}$. With the same initial angles and the decay constant $f_a$, we have also confirmed that 
the boundaries of the contours of $\,\Omega_{\rm DM} h^2 = 0.12$ are mostly insensitive to the choice of initial angles except for a slight change
in the left boundary with ${\cal R}_f > 1{}^{}$. Therefore, our results are robust against the choice of different initial angles. 

%%%%%%%%%%%%%%%%%%%%%%%%%%%%%%%%%%%%%%%%%%%%%%%%%%
%%%%%%%%%%%%%%%%%%%%%%%%%%%%%%%%%%%%%%%%%%%%%%%%%%

In summary, the dynamics of the QCD axion and ALP in the presence of the mass mixing suppresses the DM abundance,
which can expand the viable parameter space.
There are two possibilities to realize the suppression of the DM abundance:
One is the case where the level crossing takes place and ALP is produced by the adiabatic conversion of the QCD axion,
which corresponds to the case (i) in Table~\ref{tab:asymptotic_behavior}.
The other is the case where the heavy and light axions undergo a non-trivial time evolution without level crossing,
which corresponds to the case (ii) in Table~\ref{tab:asymptotic_behavior}.
In particular, even for a moderately large QCD axion decay constant which usually
leads to the overproduction of the QCD axion by the realignment mechanism, there appears a viable parameter
region where one can explain the observed DM. For this, the ALP mass must be smaller than the QCD
axion mass, while the ALP decay constant can be a few orders of magnitude smaller 
than the QCD axion. 

%%%%%%%%%%%%%%%%%%%%%%%%%%%%%%%%%%%%%%%%%%%%%%%%%%%%%%
\section{Implications for the axion search experiments}
\label{sec:axion_photons}
%%%%%%%%%%%%%%%%%%%%%%%%%%%%%%%%%%%%%%%%%%%%%%%%%%%%%%
Many axion search experiments rely on the axion coupling to photons, which is induced by loop diagrams. From Eqs. \eqref{axion_photon_coupling} and \eqref{ALP_photon_coupling}, the Lagrangian describing the interactions between the axions and photons is given as
\begin{align}
{\cal L}_{\text{axion-}\gamma\text{-}\gamma} &\,=\,
-\frac{\alpha}{8\pi}
\scalebox{1.1}{\bigg(}
{ C}_{a\gamma}\frac{a}{f_a} +
{ C}_{\varphi\gamma}\frac{\varphi}{f_\varphi} 
\scalebox{1.1}{\bigg)}
F_{\mu\nu}\widetilde{F}^{\mu\nu}\\
&\,=\, - \frac{1}{4} \scalebox{1.2}{\big(}\,g_{L\gamma\gamma} \, a_H +  g_{H\gamma\gamma} \, a_L\scalebox{1.2}{\big)} F_{\mu\nu}\widetilde{F}^{\mu\nu}~,
\end{align}
where $\,g_{H\gamma\gamma}$ and $\,g_{L\gamma\gamma}$ are the coupling constants 
in the mass eigenbasis,
\begin{eqnarray}\label{axion_photon_coupling_L}
g_{L\gamma\gamma} \,=\, 
\frac{\alpha}{2\pi f_a} 
\scalebox{1.1}{\bigg(}
{C}_{a\gamma}   \cos{\xi_0} -
{C}_{\varphi\gamma}  \frac{\sin{\xi_0}}{{\cal R}_f}
\scalebox{1.1}{\bigg)}
~,\quad
g_{H\gamma\gamma} \,=\, 
\frac{\alpha}{2\pi f_a}
\scalebox{1.1}{\bigg(}
{C}_{a\gamma}    \sin{\xi_0} +
{C}_{\varphi\gamma}  \frac{\cos{\xi_0}}{{\cal R}_f}
\scalebox{1.1}{\bigg)}
~,
\end{eqnarray}
where we have defined $\,\xi_0{}^{}\equiv\, \xi(T \to 0){}^{}$.
In our numerical study, we adopt $\,{C}_{a\gamma} {}^{}=\, {C}_{\varphi\gamma} =1 {}^{}$ as 
fiducial values. 

\begin{figure}[t]
\begin{center}
\includegraphics[scale=0.7]{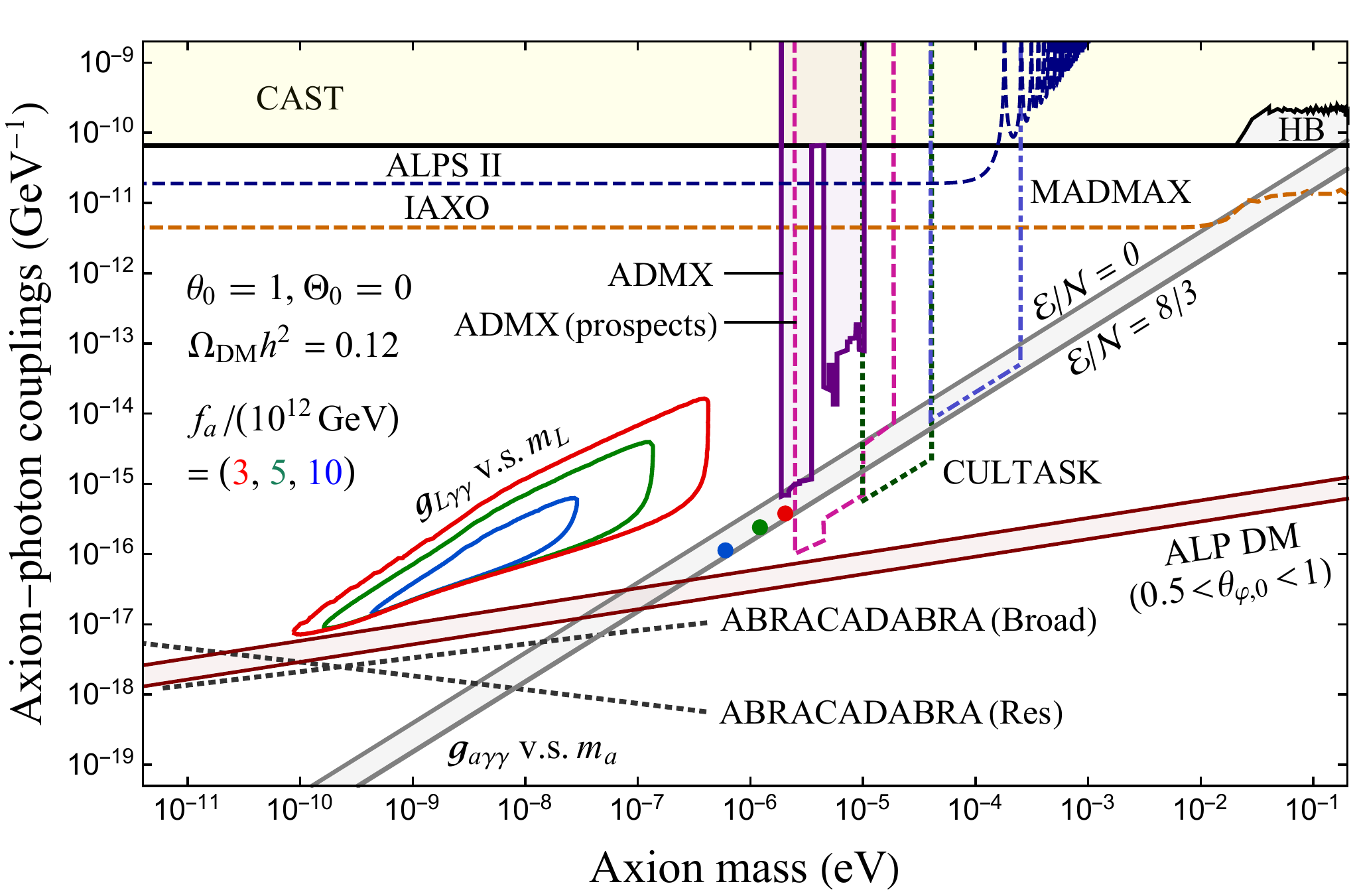}
\caption{The predicted axion-photon couplings as a function of the light axion mass for 
$f_a = (3, 5, 10)\times 10^{12}$\,GeV. 
Red, green, and blue lines represent the predictions for the mass and coupling of the light axion, and dots represent the corresponding ones of the heavy axion.
We also show experimental and astrophysical constraints on the coupling, projected sensitivities of future experiments, the QCD axion coupling with  ${\cal E}/{\cal N}=0$ and $8/3$ (see Eq.~(\ref{axion_photon_coupling})), and the region where the ALP explains DM without a mass mixing (brown diagonal band). }
\label{fig:gLr_mL}
\end{center}
\end{figure}

We map the viable regions in Fig.~\ref{fig:RfvsRm_rH} into the plane of $\,(m_L{}^{}, g_{L\gamma \gamma})$
using Eqs.~\eqref{axion_photon_coupling_L} and \eqref{mass_eigenvalues}.
The result is shown in Fig.~\ref{fig:gLr_mL}, together with various experimental and astrophysical limits
on the coupling of the QCD axion and/or ALPs with photons as a function of their mass.
We show the current constraints by ADMX  \cite{Asztalos:2003px,Asztalos:2009yp,Carosi:2013rla} and CAST~\cite{Anastassopoulos:2017ftl} and the future sensitivity regions by ADMX (prospects)~\cite{Moriond},
CULTASK~\cite{Petrakou:2017epq}, MADMAX~\cite{TheMADMAXWorkingGroup:2016hpc}, 
ABRACADABRA~\cite{Kahn:2016aff}, 
ALPS II~\cite{Bahre:2013ywa},
and IAXO~\cite{Armengaud:2014gea}. The astrophysical limit from
the studies of  the horizontal branch (HB) stars is also shown~\cite{Ayala:2014pea}.
For comparison, we show as the brown diagonal band the ALP-photon coupling $\,g_{\varphi\gamma\gamma}$ with $\,C_{\varphi\gamma}=1{}^{}$ where the ALP abundance \eqref{ALPabundance} explains DM for the initial angle between ${}^{}0.5{}^{}$ and ${}^{}1{}^{}$.  The gray solid diagonal lines show the QCD axion-photon coupling  $\,g_{a\gamma\gamma}$ with $\,{\cal E}/{\cal N} =0\,$ for the KSVZ model~\cite{Kim:1979if,Shifman:1979if} and $\,{\cal E}/{\cal N} =8/3\,$ for the 
DFSZ model~\cite{Dine:1981rt,Zhitnitsky:1980tq}. 

As expected, the coupling of the light axion with photons in our model is enhanced by a few orders of magnitude compared with the prediction of the single ALP DM shown as the brown band. Note that the sum of the light and heavy axion abundances explain DM on the contours. The heavy axion exists, but its abundance is suppressed due to the adiabatic conversion (case (i)) or 
its non-trivial evolution without level crossing (case (ii)).
The enhancement of the ALP-photon coupling is advantageous for the next generation axion-DM search experiments such as ABRACADABRA~\cite{Kahn:2016aff}. 
We also note that the three contours have an overlapping on the bottom of the triangle region. This is because
here the mixing angle is so small, $\cos \xi_0 \simeq 0$ and $\sin \xi_0 \simeq 1{}^{}$,  
that $|g_{L\gamma\gamma}| \simeq \alpha/(2\pi f_\varphi)$ is independent of $f_a$. (The contours in Fig.~\ref{fig:RfvsRm_rH} were for different values of $f_a$).
The smallness of the mixing angle also implies that the couplings of the light axion with nucleons
are suppressed, which makes it difficult to detect DM in experiments that make use of nucleon couplings.

Comparing with the right panel of Fig.~\ref{fig:RfvsRm},
we see that the relative fraction of the heavy axion in the total DM abundance becomes non-negligible 
in the upper-left region of the triangle-shaped predictions for the mass and coupling of the light axion shown in Fig.~\ref{fig:gLr_mL}.
In this region, the contribution of the heavy axion to the total DM abundance can be 
as large as $\mathcal{O}(10)\,\%$
and the predictions for its mass and coupling
are represented as dots in Fig.~\ref{fig:gLr_mL}.
Although the values of $g_{H\gamma\gamma}$ and $m_H$ for $f_a = (3,5,10)\times 10^{12}\,\mathrm{GeV}$ plotted in Fig.~\ref{fig:gLr_mL}
are out of the experimental sensitivities, we expect that they would be within the sensitivities of future DM experiments such as 
ADMX, CULTASK, and MADMAX, if we consider slightly smaller values of $f_a \sim \mathcal{O}(10^{11})\,\mathrm{GeV}$.
Therefore, there is a possibility to detect both the heavy and light axions in future axion search experiments if $f_a \sim \mathcal{O}(10^{11})\,\mathrm{GeV}$ (or smaller)
and the contribution of the heavy axion to the total DM abundance is sizable.
On the other hand, if the relative fraction of the heavy axion in the total DM abundance is negligibly small
(the lower-right region of the triangle-shaped boundaries shown in Fig.~\ref{fig:gLr_mL}),
it becomes impossible to detect the heavy axion even in future experiments.

%%%%%%%%%%%%%%%%%%%%%%%%%%%%%%%%%%%%%%%%%%%%%%%%%%
\section{Discussion and Conclusions}
\label{sec:conclusion}
%%%%%%%%%%%%%%%%%%%%%%%%%%%%%%%%%%%%%%%%%%%%%%%%%%

In this paper, we have considered cosmology of the QCD axion and ALP DM and
studied the scenario where they
%the QCD axion and ALP 
have a mass mixing, focusing 
on the adiabatic conversion between them. Throughout the paper we have assumed that the QCD axion and ALP are spatially homogeneous, which
enables us to analyze their dynamics in terms of the system of multiple harmonic oscillators. 
Before concluding our study, let us discuss validity of this assumption and enumerate other possibilities to clarify
difference between our case and the case without the mass mixing as well as the limitation of our results.

Strictly speaking, both the QCD axion and ALP are not exactly homogeneous, and they can acquire quantum fluctuations during inflation, 
which induce isocurvature perturbations. There is a stringent limit on the isocurvature perturbations~\cite{Ade:2015lrj}. The isocurvature perturbation of the QCD axion
is usually given by the ratio of the Hubble parameter during inflation to the decay constant, and one way to satisfy the isocurvature limit is to consider low-scale inflation. In our scenario, 
we can explain the DM abundance for a larger $f_a$ compared to the case without the mass mixing, and
the isocurvature constraint on the inflation scale can be relaxed. 
In the case without the mass mixing, the ALP coupling to photons can be enhanced by
considering the anharmonic effect which enhances the ALP abundance, thereby the small value of the ALP decay constant is required in order to satisfy the relic abundance of DM.  However, it simultaneously enhances the isocurvature perturbation as well as its non-Gaussianity~\cite{Kobayashi:2013nva}.

Another way to avoid the isocurvature limit is to assume that the corresponding global U(1) symmetries are restored
during or after inflation. In this case, topological defects such as cosmic strings and domain walls are formed after the symmetries get spontaneously broken. In the presence of multiple axions with the mass mixings, the strings and walls are considered to form a complicated network~\cite{Higaki:2016jjh,Long:2018nsl}. Such a structure naturally appears in the QCD axion with a clockwork (or alignment) structure~\cite{Higaki:2015jag,Higaki:2016yqk}. In the present model, however, we have only two axions, and the two kinds of the cosmic strings will be attached to each other once the heavy axion starts to oscillate and domain walls are formed. If $m_\varphi \gg m_a{}^{}$, they behave as the ordinary cosmic strings for the QCD axion, and the subsequent evolution is similar to the usual scenario without the mass mixing. On the other hand, if $m_\varphi < m_a{}^{}$, one needs to consider both the string dynamics and the adiabatic conversion during and after the QCD phase transition, which is beyond the scope of this paper. 

Finally, let us summarize the results obtained in this paper.
Compared to the previous work~\cite{Kitajima:2014xla} in which the suppression of the QCD axion abundance due to the mass mixing was investigated, 
our main novelties are the refinement of the conditions for the adiabatic conversion, the identification of the parameter region in which the light and/or heavy axions can explain DM,
and the discussion on the experimental implications.
We have fully explored the parameter space to find the condition for the level crossing to take place and refined the adiabatic condition for the comoving axion numbers, $N_H$ and $N_L{}^{}$, to be separately conserved. We have found that the adiabatic condition involves the ratio of the decay constants, $f_\varphi/f_a{}^{}$, and numerically confirmed that our refined condition indeed provides the boundary of the viable parameter region where the DM is explained by the light and heavy axions. 
From the numerical results, we also have shown that the effect of the mass mixing suppresses the axion abundance, opening up a new parameter space,
which was impossible in the case without mass mixing unless the initial angles are fine-tuned or temperature dependence of the ALP mass is introduced.
Here, the suppression is caused by the adiabatic conversion or modified mass eigenvalues due to the existence of the mass mixing.
Interestingly, the light axion has a larger coupling to photons in the new parameter region compared to the
conventional scenario without the mixing for the same mass range,
and such a parameter region will be within reach of the future axion search experiments.

\section*{Acknowledgments}
We thank Georg Raffelt and Andreas Ringwald for useful comments. 
F.T. thanks the hospitality of MIT Center for Theoretical Physics where a part of this work was done. 
K.S. acknowledges partial support by the Deutsche Forschungsgemeinschaft through Grant No.\ EXC 153 (Excellence
Cluster ``Universe'') and Grant No.\ SFB 1258 (Collaborative Research
Center ``Neutrinos, Dark Matter, Messengers'') as well as by the
European Union through Grant No.\ H2020-MSCA-ITN-2015/674896 (Innovative Training Network ``Elusives'').
This work is partially supported by JSPS KAKENHI Grant Numbers JP15H05889 (F.T.), 
JP15K21733 (F.T.), JP17H02878 (F.T.), and JP17H02875 (F.T.), Leading Young Researcher Overseas
Visit Program at Tohoku University (F.T.), and by World Premier International Research Center
 Initiative (WPI Initiative), MEXT, Japan (F.T.).

\end{document}